\documentclass[11pt]{article}
\usepackage{epsfig}
\usepackage{amsmath,amsfonts,amssymb} 
\usepackage{color}
\usepackage{cite}
\usepackage{slashed}

\newcommand{\unit}{\leavevmode\hbox{\small1\kern-3.6pt\normalsize1}}

\def\chsnsn{C_{\higgsi\tilde\nu\tilde\nu}}

\def\csnnneui{C_{\tilde\nu N\neuti}}

\def\vevs{v_s}

\def\ln{{\lambda_N}}
\def\aln{{A_{\lambda_N}}}
\def\mn{{m_{\tilde{N}}}}
\def\ayn{A_{y_N}}
\def\yn{y_N}

\def\charg{{\tilde\chi}^\pm}

\def\neuti{{{\tilde\chi}_{i}}}

\def\neutl{{{\tilde\chi}^0_{1}}}
\def\neut{{\tilde\chi}^0}

\def\neutmass{{m_{{\tilde\chi}_{1}^0}}}

\def\snr{{\tilde N}}
\def\snl{{\tilde\nu_L}}

\def\snmassr{{m_{\tilde N_1}}}
\def\snmassrsq{{m^2_{\tilde N_1}}}

\def\sncompl{{N^{\tilde\nu}_{i \tilde\nu_L}}}
\def\sncompr{{N^{\tilde\nu}_{i \tilde N}}}

\def\sncompl{{N^{\tilde\nu}_{i L}}}
\def\sncompr{{N^{\tilde\nu}_{i R}}}
\def\snlcompr{{N^{\tilde\nu}_{1 R}}}
\def\snrcompl{{N^{\tilde\nu}_{2 L}}}

\def\lhn{\nu_L}

\def\rhn{{N}}
\def\rhnmass{{M_{N}}}

\def\ncompl{{N^{\nu}_{i L}}}
\def\ncompr{{N^{\nu}_{i R}}}
\def\nmixl{{N^{\nu}_{1 R}}}
\def\nmixr{{N^{\nu}_{2 L}}}

\def\hsm{h_{SM}^0}

\def\higgsi{{H_i^0}}

\def\higgsl{{H_1^0}}

\def\hmasssm{m_{\hsm}}

\def\phiggsl{{A_1^0}}

\newcommand{\higgs}[1]{H^{0}_{#1}}

\def\met{\slash\hspace*{-1.5ex}E_{T}}

\newcommand{\bsg}{b\to s\gamma}

\newcommand{\asusy}{a^{\rm SUSY}_\mu}

\def\lsim{\raise0.3ex\hbox{$\;<$\kern-0.75em\raise-1.1ex\hbox{$\sim\;$}}}
\def\gsim{\raise0.3ex\hbox{$\;>$\kern-0.75em\raise-1.1ex\hbox{$\sim\;$}}}

\allowdisplaybreaks

\begin{document}

\thispagestyle{empty}
\begin{flushright}
  IFT-UAM/CSIC-13-126\\
  FTUAM 13/36\\
  HGU-CAP-027\\

  \vspace*{2.mm} November 27, 2013
\end{flushright}
\def\thefootnote{\fnsymbol{footnote}}
\begin{center}
  {\Large \textbf{Displaced vertices and long-lived charged particles in the NMSSM with right-handed sneutrinos
  } }  
  
  \vspace{0.5cm}
  David G.~Cerde\~no${}^{1,2,}$\footnote{davidg.cerdeno@uam.es},
  V\'ictor Mart\'in-Lozano${}^{1,2,}$\footnote{victor.martinlozano@uam.es},
  Osamu Seto${}^{3,}$\footnote{seto@physics.umn.edu} \\[0.2cm] 
    
  {\small \textit{ ${}^1$ 
      Instituto de F\'{\i}sica Te\'{o}rica
      UAM/CSIC, Universidad Aut\'{o}noma de Madrid,\\ 
      Cantoblanco, E-28049, Madrid, Spain\\[0pt] 
      ${}^2$ 
      Departamento de F\'{\i}sica Te\'{o}rica,
      Universidad Aut\'{o}noma de Madrid, 
      Cantoblanco, E-28049, Madrid, Spain\\[0pt] 
      ${}^3$ 
      Department of Life Science and Technology,
      Hokkai-Gakuen University,\\[0pt]
      Asahimachi, 062-8605, Sapporo, Japan  }}
  
\vspace*{0.7cm}

\begin{abstract}
We study LHC signatures of displaced vertices and long-lived charged particles within the context of the Next-to-Minimal Supersymmetric Standard Model with right-handed (RH) sneutrinos.
In this construction the RH neutrino can be produced directly from Higgs decays or in association with a RH sneutrino when the latter is the lightest supersymmetric particle. 
The RH neutrino is generally 
long-lived, since its decay width is proportional to the neutrino Yukawa, a parameter which is predicted to be small.
The RH neutrino late decay can therefore give rise to displaced vertices at the LHC, which can be identified through 
the decay products, which involve two leptons ($2\ell + \met$) or a lepton with two jets ($\ell j j $). 
We simulate this signal for the current LHC configuration (a centre of mass of 8~TeV and an integrated luminosity of $\mathcal{L}=20$ fb$^{-1}$), and a future one (13~TeV and $\mathcal{L}=100$ fb$^{-1}$).
We show that a region of the parameter space of this model can be probed and that the RH neutrino mass can be reconstructed from the end-point of the two-lepton invariant mass distribution or the central value of the mass distribution for two jets plus one lepton. 
Another exotic signature of this construction is the production of a long-lived stau. If the stau is the next-to-lightest supersymmetric particle, it can decay through diagrams involving the small neutrino Yukawa, and would escape the detector leaving a characteristic trail. 
We also simulate this signal for various benchmark points and show that the model can be within the reach of the future run of the LHC. 
\end{abstract}

\end{center}

\newpage
	

\renewcommand{\thefootnote}{\arabic{footnote}}
\setcounter{footnote}{0}

\section{Introduction}

The Large Hadron Collider (LHC) is probing the nature of Physics beyond the Standard Model (SM) at an unprecedented energy scale, having reached a centre of mass energy of 8~TeV and an integrated luminosity of $\mathcal{L}\approx20$ fb$^{-1}$ at the end of its first years of operation. 
An indisputable achievement in this first period has been the observation of a Higgs boson with a mass
in the $2\,\sigma$ range $124-126.8$~GeV ($124.5-126.9$~GeV)
by ATLAS (CMS)~\cite{Higgs,CMS:2012gu,ATLAS:2013mma,CMS:yva}
with very similar properties to those predicted by the SM.
On the other hand, unsuccessful searches for exotic signals have allowed us to set stringent constraints on models for new physics. Such is the case of Supersymmetry (SUSY), for which 
a lower bound on the mass of the gluino and squarks can be derived. For example, in simplified scenarios such as a constrained version of the Minimal Supersymmetric Standard Model (MSSM), where mass parameters are assumed to be universal, one  obtains $m_{\tilde q},\ m_{\tilde g}\gsim 1.2$~TeV (and even $m_{\tilde g}\gsim1.8$~TeV if $m_{\tilde q}=m_{\tilde g}$)
\cite{susy-latest,CMS-susy-multijet,TheATLAScollaboration:2013fha,TheATLAScollaboration:2013uha,Chatrchyan:2013iqa,Chatrchyan:2013xna}.
There are a number of signatures which are generic to most SUSY models. This is, e.g, the case of multi-lepton/jet signals with missing energy associated to the lightest supersymmetric particle (LSP) if the latter is neutral and stable (typically, the lightest neutralino). 
There are other signatures which are more exotic but which may be used to discriminate among different scenarios. This is, for example, the case of displaced vertices (due to late decaying neutral particles) and long-lived charged particles (which leave a characteristic track in the detector).

In this work we investigate the production of displaced vertices and long-lived charged particles within the context of the Next-to-Minimal Supersymmetric Standard Model (NMSSM) with a right-handed (RH) neutrino/sneutrino \cite{Cerdeno:2008ep,Cerdeno:2009dv}. 
This construction features two singlet superfields, as in 
Refs.\,\cite{ko99,pilaftsis}. A singlet superfield, $S$, is the usual NMSSM scalar Higgs which 
addresses the $\mu$ problem  \cite{Kim:1983dt}
and provides extra Higgs and neutralino states, while an extra singlet
superfield, $N$, accounts for RH neutrino and sneutrino
states. 
In this construction, the RH neutrino mass is generated with the electroweak symmetry breaking mechanism through the new coupling $S N N$.
Due to the non-vanishing vacuum expectation value (VEV) of the singlet Higgs, an effective 
Majorana mass for the RH neutrino is generated which is of the order of the electroweak
scale, in the same way as
the effective $\mu$ term \cite{Cerdeno:2008ep}.
This implies a low scale see-saw mechanism for neutrino mass generation, which entails a small Yukawa coupling, $\yn\sim10^{-6}$, thus leading to a tiny mixing between RH and left-handed (LH) fields. 
An interesting feature of this construction is that the RH sneutrino can be a viable candidate for weakly interacting massive particle (WIMP) dark matter  \cite{Cerdeno:2008ep,Cerdeno:2009dv,Cerdeno:2011qv} if it is the LSP.

The smallness of the neutrino Yukawa has very interesting implications for LHC phenomenology.
On the one hand, the RH neutrino, which decays into SM particles through the mixing with the LH neutrino, can be long-lived enough to give rise to a displaced vertex in the inner detector that can be observed through the emitted leptons or jets.
The RH neutrino 
can appear directly in decays of the Higgs boson or at the end of a supersymmetric decay chain in association with a RH sneutrino, when the latter is the LSP. Since the RH neutrino is relatively easy to produce, displaced vertices can be a characteristic signature of this model.
In this work we study this possibility in detail.
We show that the late RH neutrino decay can be observed as two leptons ($2\ell + \met$) or a lepton with two jets ($\ell j j $) events.
We carry out a Monte Carlo simulation to determine the number of events expected at the current and future LHC configurations for various representative benchmark points. 
Through the study of the resulting two-lepton ($m_{\ell\ell}$) and two-jets one lepton ($m_{\ell j j }$) invariant mass distributions, we argue that the end-point in $m_{\ell\ell}$ and the peak in $m_{\ell j j }$ can give valuable information with which the mass of the RH neutrino can be reconstructed.

On the other hand, the decay of the next-to-lightest supersymmetric particle (NLSP) into a RH sneutrino LSP can also be suppressed by the neutrino Yukawa in certain regions of the parameter space. This is, for example, the case of the lighter stau which, being a charged particle, would leave a characteristic track after crossing the whole detector. In this paper we also investigate this possibility. We consider two benchmark points with a long-lived stau and simulate their production in the current and future LHC configurations.

The paper is organised as follows. In Section\,\ref{sec:rhsn} we present the main features of the NMSSM with RH neutrinos/sneutrinos and introduce our notation. We also include the most recent LHC constraints on the Higgs sector, with especial attention to the bounds on invisible and exotic Higgs decays, and determine the relevant areas of the parameter space. 
In Section\,\ref{sec:displaced} we investigate the displaced vertices that can be originated by the late decay of RH neutrinos. The case of long-lived staus is studied in Section\,\ref{sec:stau}. 
Finally, the conclusions are presented in Section\,\ref{sec:conclusions}.

\section{The NMSSM with right-handed neutrino/sneutrino}
\label{sec:rhsn}

The NMSSM with RH neutrino and sneutrino states was introduced in Refs.\cite{Cerdeno:2008ep,Cerdeno:2009dv}. It was there shown that the RH sneutrino can be the LSP and a viable candidate for dark matter within the category of WIMPs, since the correct relic abundance can be obtained in wide regions of the parameter space, including the possibility that the RH sneutrino is very light \cite{Cerdeno:2011qv}.

The superpotential of this model reads
\begin{equation}
	W = W_{\rm NMSSM} + \lambda_N S N N + y_N L \cdot H_2 N,   
	\label{superpotential}
\end{equation}
where flavour indices are omitted and the dot denotes the $SU(2)_L$
antisymmetric product. The Lagrangian contains
new soft-supersymmetry breaking parameters as follows
\begin{equation}
	-{\cal L} = -{\cal L}_{\rm NMSSM}
	+ \mn^2 |\tilde{N}|^2  + \left( 
	\ln\aln S \tilde{N}^2 + \yn\ayn \tilde{L} H_2 \tilde{N}+ {\rm H.c.}  \right) .
	\label{lagrangian_couplings}
\end{equation}
In total,  five new free parameters are included, namely a soft sneutrino mass $\mn$, two Yukawa couplings $\ln$ and $\yn$, and two trilinear parameters $\aln$ and $\ayn$.
After radiative Electroweak symmetry-breaking takes place the Higgs fields take 
non-vanishing VEVs. In particular the VEV of the singlet, $\vevs$, triggers an effective $\mu$ parameter which provides an elegant solution to the $\mu$ problem of the MSSM. At the same time, an effective Majorana mass is generated for the RH neutrino,  
\begin{equation}
	\rhnmass = 2 \ln v_s\ ,
\label{RighthandedNeutrinoMass}
\end{equation}
which is therefore of the order of the electroweak scale.

The neutrino mass matrix can then be written in terms of the above quantities as
\begin{eqnarray}
	M_\nu=
	\left(
		\begin{array}{cc}
		0 & \mathbf{y}_N v_2 \\
		\mathbf{y}^T_N v_2 & 2\mathbf{\lambda}_N v_s
	\end{array}
	\right)=
	\left(
		\begin{array}{cc}
		0 & M_D \\
		M_D^T& M_N
	\end{array}
	\right) .
	\label{eq:nmatrix}
\end{eqnarray}
In general $M_D$ is a $3\times k$ matrix and $M_N$ is a $k\times k$ matrix, where $k$ is the number of RH neutrinos. In this work, for simplicity, we consider only one RH neutrino with equal mixings with the three left-handed neutrinos\footnote{The properties of a general construction, with three RH neutrinos, would be affected by the specific texture of the Yukawa matrix. Although this would have a profound effect on the resulting neutrino phenomenology, the presence of displaced vertices would be treated in a similar fashion to the analysis in this work.}.
In the limit where the neutrino Yukawa is small, the diagonalization of the above mass matrix yields two eigenstates which are almost approximately pure gauge eigenstates. The lightest of these would correspond to ordinary left-handed neutrinos, $\nu_1=\nu_L$, whereas the heavier one is a pure RH neutrino $\nu_2=N$, with masses as follows,
\begin{equation}
	m_{\nu_L}= \frac{\yn^2 v_2^2}{2\ln\vevs}\,, \quad\quad \rhnmass=2\ln\vevs\,.
\end{equation}
Notice that in order to reproduce the smallness of the left-handed neutrinos the value of $\yn$ has to be small, of the order of the electron Yukawa, $\yn\sim10^{-6}$, typical of a low-scale see-saw mechanism. As we will see in the next section, the smallness of this parameter is responsible for the presence of displaced vertices or long-lived charged particles.

We will express neutrino mass eigenstates in terms of the mixing matrix, $N^\nu$ as follows,
\begin{equation}
	{\nu}_i = \ncompl \lhn + \ncompr \rhn\,,
\end{equation}
and identify $\nu_1\approx\lhn+\nmixl\rhn$ and $\nu_2\approx\nmixr\lhn+\rhn$. The mixing between LH and RH neutrino mass eigenstates, as obtained from the diagonalization of eq.\,(\ref{eq:nmatrix}), is proportional to $\yn$ and therefore small, $\nmixl=\nmixr=\frac{\yn v_2}{2\ln\vevs}$.

Regarding the sneutrino sector, the mass eigenstates are also a linear superposition of the LH and RH gauge eigenstates, $\snl$ and $\snr$, respectively. We can use a similar description in terms of the mixing matrix $N^{\tilde \nu}$ as follows,
\begin{equation}
	\tilde{\nu}_i = \sncompl \tilde{\nu}_L + \sncompr \tilde{N}\,.
\end{equation}
As in the case of the neutrinos, the left-right mixing terms are proportional to $\yn$ (the complete expression can be found in Ref.\,\cite{Cerdeno:2009dv}) and are therefore very small. For this reason the mass eigenstates are almost pure LH or RH fields, $\tilde\nu_1\approx\snrcompl\tilde{\nu}_L+\snr\approx\snr_1$ and $\tilde \nu_2\approx\tilde{\nu}_L+\snrcompl\snr\approx\tilde\nu_L$,
with $\snrcompl,\,\snlcompr={\cal O}(\yn)$. Notice that in this case we identify the lightest eigenstate with the lighter RH sneutrino, $\snr_1$. In terms of the rest of the parameters the lighter RH sneutrino mass reads
\begin{equation}
	\snmassrsq= m_{\tilde{N}}^2 +|2\lambda_N \vevs|^2  + |y_N v_2|^2 
	\pm 2 \lambda_N \left( A_{\lambda_N} \vevs+
	(\kappa \vevs^2-\lambda v_1 v_2 )^{\dagger} \right) ,
	\label{eq:snmass}
\end{equation}
where the sign in front of $2\ln$ is chosen opposite to the sign of $2\ln\left(A_{\lambda_N} \vevs+(\kappa \vevs^2-\lambda v_1 v_2 )^{\dagger} \right) $.

In this construction, the on-shell production of RH neutrinos can lead to the occurrence of displaced vertices. 
Moreover, if the NLSP is the lighter stau, it can also behave as a long-lived charged particle if produced on-shell.
In both cases the lifetime of the corresponding particle is a function of the neutrino Yukawa, $\yn$, as we will see in the next section, and since $\yn\sim10^{-6}$ particles tend to be long-lived.

Throughout the paper we consider input parameters defined at the electroweak scale, so no running is performed. The supersymmetric spectrum and Higgs phenomenology is computed using {\tt NMSSMTools} \cite{Ellwanger:2004xm, Ellwanger:2005dv, Ellwanger:2006rn}, which we have modified to incorporate the RH neutrino and sneutrino sector. We also include a condition on the stability of the corresponding vacuum following the analysis of Ref.~\cite{Kanehata:2011ei}.
The decay width for the RH neutrino has been calculated using {\tt CalcHEP 3.4}~\cite{Belyaev:2012qa}. 
We incorporate the most recent experimental constraints on the masses of supersymmetric particles, as well as on low-energy observables (which are also computed using {\tt NMSSMTools}). 
In particular, we consider the recent measurement of the branching ratio of the $B_s\to \mu^+\mu^-$ process
by the LHCb \cite{Aaij:2013aka} and CMS \cite{Chatrchyan:2013bka} collaboration, which implies $1.5\times 10^{-9}< {\rm BR}(B_s\to \mu^+\mu^-)< 4.3\times 10^{-9}$ at
95\% CL. 
Also, for the $b\to s\gamma$ decay, we require the 2$\sigma$ range
$2.89\times 10^{-4}< {\rm BR}(b\to s\gamma)< 4.21\times
10^{-4}$, where
theoretical and experimental uncertainties have been added in quadrature \cite{Ciuchini:1998xy,D'Ambrosio:2002ex,Misiak:2006zs,
Misiak:2006ab,Amhis:2012bh}.
We also impose the constraint on the branching ratio of the $B^+\to\tau^+\nu_\tau$ decay at $2\sigma$, $0.85\times 10^{-4}< {\rm BR}(B^+\to \tau^+ \nu_\tau)< 2.89\times 10^{-4}$\cite{Lees:2012ju}.
Regarding the supersymmetric contribution to the muon anomalous magnetic moment, $\asusy$,  
experimental data using $e^+e^-$ suggest that there is a deviation from the SM value
\cite{Bennett:2006fi,Jegerlehner:2009ry,Gray:2010fp,Davier:2010nc,Hagiwara:2011af}. However, if tau data is used,
this discrepancy is smaller \cite{Davier:2010nc}. In our analysis we compute this quantity but do not impose any constraints on it. 
Following the recent observations, we demand the presence of a Higgs boson with a mass of $126$~GeV and SM-like couplings \cite{Higgs,CMS:2012gu}. 
Finally, some analysis suggest the existence of a second singlet-like Higgs boson with a mass around $98$~GeV \cite{Barate:2003sz,Schael:2006cr,Drees:2012fb}, a possibility that we also consider in one example.

Table \ref{tab:S-12} shows the input parameters for three NMSSM scenarios, labelled S1, S2 and S3, that will be used in this paper and that pass all the constraints mentioned above. 
We also indicate the RH sneutrino relic density and spin-independent scattering cross section off nucleons
\footnote{We only give these quantities for information, since we have not applied dark matter constraints in this work. Most of the points have a relic density very close to the value obtained by Planck data and
a value of $\sigma^{SI}$ that is just above or below the current upper bound obtained by the LUX, XENON100 and SuperCDMS direct detection experiments \cite{Akerib:2013tjd,Aprile:2012nq,Agnese:2014aze}. 
Direct detection limits are more important for points with light RH sneutrino, such as S1a and S1b, and some of us will reanalyse the viability of light RH-sneutrinos in the light of these bounds \cite{sandra}.
}, $\sigma^{SI}$. 
Part of the resulting supersymmetric spectrum (corresponding to the Higgs, stau and neutralino/chargino sectors) is shown, together with the corresponding values for some low-energy observables. In scenario S1 the SM Higgs is the second-lightest one, $\hsm=H^0_2$, whereas in scenarios S2 and S3 it is the lightest one, $\hsm=H^0_1$.

Since we have chosen small values of $\tan\beta$, the value of BR(${B_S\to\mu^+\mu^-}$) is very close to the SM value and this constraint is not very important in ours scan. On the other hand, BR($\bsg$) has a more serious impact on the NMSSM parameter space (see e.g., Ref.\,\cite{Cerdeno:2007sn}). Finally, in the low $\tan\beta$ regime the contribution to the muon anomalous magnetic moment is not sufficient to account for the deviation observed in $e^+e^-$ data.

In Refs.\,\cite{Cerdeno:2009dv,Cerdeno:2011qv} we showed that the RH sneutrino relic density can be adjusted by playing with the free parameters $\ln$, $\aln$ and $\mn$ without significantly affecting the NMSSM phenomenology. For this reason, in this analysis we do not impose any constraint on the relic abundance of the RH sneutrino.

\begin{table}[!h]
\begin{center}
\hspace*{-0.7cm}\scalebox{0.86}{
\begin{tabular}{|c|c|c|c|}
\hline
{\bf  \begin{minipage}{2cm}\begin{center}\rule{0pt}{4.5ex}Scenarios\\ \quad \end{center}\end{minipage}}
&{\bf S1}&{\bf S2}&{\bf S3}
\\
\hline
\hline
$\tan \beta$&2.0&2.5&2.7\\
$M_1,\ M_2,\ M_3$&500, 650, 1950 & 300, 600, 1800 & 345, 575, 2500 \\
$m_{L,E}$&300 &250 & 1000,350 \\
$m_{{Q,U,D}_{1,2}}$&2000 &2000 &2000 \\
$m_{{Q,U,D}_{3}}$&1500 &2000 &2000\\
$A_E$& -1000 &-1000 &750\\
$A_{U,D}$& 2000 &2300 &2550\\
$\mu$& 152 &180 &595\\
$\lambda,\ \kappa$& 0.50, 0.27& 0.60, 0.40& 0.58, 0.34\\
$A_\lambda,\ A_{\kappa}$&283, -220 &265, -50 &1189, -225 \\
\hline
\hline
$m_{H^0_1},\,\,m_{H^0_2},\,\,m_{H^0_3}$&99.5,  125.8,  358.6&125.7,  225.7,  446.2&125.8,  656.9,  1650.5\\
$m_{A_1},\,\,m_{A_2}$&254.1, 348.9& 181.0, 432.8& 501.5, 1644.9\\
$m_{\tilde{\chi}^0_1},\,\,m_{\tilde{\chi}^0_2},\,\,m_{\tilde{\chi}^0_3}$&127.0, 176.3, 200.1& 147.3, 206.9, 277.9& 335.7, 528.9, 611.2\\
$m_{\tilde{\chi}^0_4},\,\,m_{\tilde{\chi}^0_5}$&492.3, 674.3& 306.8, 627.6& 665.5, 740.7\\
$m_{\tilde{\chi}^\pm_1},\,\,m_{\tilde{\chi}^\pm_2}$&144.9, 674.1& 173.6, 627.5& 530.6, 676.5\\
$m_{\tilde{\tau}_1},\,\,m_{\tilde{\tau}_2}$&290.5, 312.9& 245.9, 259.5& 352.0, 1000.8\\
\hline
\hline
BR($b\to s\gamma$)&
$4.244^{+0.436}_{-0.631}\,\times 10^{-4}$&$3.984^{+0.381}_{-0.578}\,\times 10^{-4}$&$3.307^{+0.256}_{-0.456}\,\times 10^{-4}$
\\
BR($B_s\to \mu^+\mu^-$)&
$3.676^{+2.567}_{-1.891}\,\times 10^{-9}$&$3.677^{+2.568}_{-1.892}\,\times 10^{-9}$&$3.677^{+2.568}_{-1.892}\,\times 10^{-9}$
\\
BR($B^+\to \tau^+\nu_{\tau}$)&
$1.316^{+1.316}_{-0.748}\,\times 10^{-4}$&$1.316^{+1.316}_{-0.748}\,\times 10^{-4}$&$1.318^{+1.318}_{-0.749}\,\times 10^{-4}$
\\
$a_{\mu}^{SUSY}$&
$2.717^{+2.906}_{-2.528}\,\times 10^{-10}$&$4.592^{+2.938}_{-2.938}\,\times 10^{-10}$&$5.142^{+2.818}_{-2.637}\,\times 10^{-10}$
\\
\hline
\hline
{\bf \begin{minipage}{2cm}\begin{center}\rule{0pt}{3ex}Benchmark\\ Points\\ \quad \end{center}\end{minipage}}
&{\bf S1a}\quad\quad {\bf S1b}\quad \quad {\bf S1c}&{\bf S2a}\quad\quad  {\bf S2b}\quad\quad {\bf S2c}&{\bf S3a}\quad\quad{\bf S3b}
\\
\hline
$\begin{array}{c}\ln\\ \mn\\ \aln\\ \yn \end{array}$
&
$\begin{array}{ccc} 0.165&0.091&0.017\\ 92.2 &128.9 &80.6\\-250&-250&-250\\10^{-7}&10^{-6}&10^{-5} \end{array}$
&
$\begin{array}{ccc}0.067&0.033&0.017\\ 68.5&130.9&42.5\\ -150&-150&-150\\ 10^{-6}&10^{-6}&10^{-5} \end{array}$
&
$\begin{array}{cc} 0.083 & 0.151\\ 190.7&179.2\\ -500&-750\\ 10^{-7}&10^{-7} \end{array}$
\\
\hline
\hline
$\begin{array}{c}\snmassr \\ \rhnmass \end{array}$
&
$\begin{array}{ccc} 20\quad&100\quad&70\\ 100\quad &55\quad &10 \end{array}$
&
$\begin{array}{ccc}70\quad&130\quad&40\\ 40\quad&20\quad&10 \end{array}$
&
$\begin{array}{cc} 200\quad & 65\\ 170\quad&310 \end{array}$
\\
\hline
$\begin{array}{c}\Omega_{\tilde{N}_1} h^2 \\  \sigma^{\rm SI}\times 10^{7} \end{array}$
&
$\begin{array}{ccc} 0.356&0.155&21.2\\2.4  &2.0 &6.9\times 10^{-4} \end{array}$
&
$\begin{array}{ccc}0.684&0.838&65.6\\ 7.4\times 10^{-3}&5.4\times 10^{-4}&1.4\times 10^{-3} \end{array}$
&
$\begin{array}{cc} 0.729 & 0.047\\ 1.1\times 10^{-3}&3.3 \times 10^{-2}\end{array}$
\\
\hline
\end{tabular}}
\end{center}
\vspace*{-0.5cm}
 \caption{\small Input parameters of the NMSSM at the electroweak scale that define the three scenarios S1, S2 and S3 used in this work. The resulting masses of the scalar and pseudoscalar Higgses are indicated, together with the neutralinos, charginos, and the lighter stau, as well as the values of some low energy observables with the corresponding theoretical error.
For each scenario, a number of representative benchmark points are defined by the corresponding values of the soft RH sneutrino mass, $\mn$, soft trilinear parameter, $\aln$, coupling $\ln$, and Yukawa coupling $\yn$. We also indicate the RH sneutrino mass, $\snmassr$ and RH neutrino mass $\rhnmass$, as well as the RH sneutrino relic density and spin-independent scattering cross section off nucleons (in pb).
All the masses are given in GeV.
}
      \label{tab:S-12}
\end{table}

\subsection{Constraints on the Higgs invisible decay width}
\label{sec:invisible}

The recently discovered Higgs particle at the LHC has a mass of $126$~GeV and SM-like branching ratios \cite{Higgs, CMS:2012gu}. Within the NMSSM a scalar Higgs with these properties can be obtained in wide regions of the parameter space \cite{Ellwanger:2011sk,Hall:2011aa,Ellwanger:2011aa,Gunion:2012zd,Arvanitaki:2011ck,King:2012is,Kang:2012sy,Cao:2012fz, Ellwanger:2012ke,Benbrik:2012rm,Gunion:2012gc,Cao:2012yn,Belanger:2012tt,Kowalska:2012gs,King:2012tr}. 
In fact, the presence of an extra scalar Higgs field induces new contributions to the Higgs mass from the $\lambda S H_uH_d$ term in the superpotential, which allows to get a fairly heavy Higgs boson while reducing the fine-tuning with respect to the situation in the MSSM. 
The Higgs sector of the NMSSM is very rich, and the presence of a lighter scalar Higgs is also allowed, provided that it is mostly singlet-like. 
All these features are still valid in our construction, however, when implementing constraints on the resulting Higgs phenomenology one has to be aware that the presence of light RH neutrinos or sneutrinos can contribute significantly to the invisible decay width of the scalar Higgses  \cite{Cerdeno:2011qv}. 
For the reduced
signal strength
of the Higgs to di-photon mode, $R_{\gamma\gamma}$, we
use $0.23\leq R_{\gamma\gamma}\leq 1.31$, the latest CMS results at
2$\sigma$~\cite{CMS:yva}\footnote{For ATLAS the same
limit including all systematics is $0.95\leq R_{\gamma\gamma}\leq 2.55$ \cite{ATLAS:2013oma,ATLAS:2013wla}.}.
The remaining reduced signal strengths are also constrained according
to the CMS results of Ref.~\cite{CMS:yva} (see Refs.~\cite{ATLAS:2013mma,ATLAS:2013wla} for the equivalent ATLAS results).
Notice that these measurements indirectly entail a
strong bound on the invisible and non-standard
decay modes of the SM-like Higgs boson~\cite{ATLAS:2013pma,
Espinosa:2012vu,Belanger:2013kya,Falkowski:2013dza,Giardino:2013bma,Ellis:2013lra,Djouadi:2013qya,Belanger:2013xza}, which
in our case affects the decay modes
$\hsm\to\higgsl\higgsl$, $\hsm\to\phiggsl\phiggsl$, $\hsm\to\neut_i\neut_i$, and especially, $\hsm\to\rhn\rhn$ and $\hsm\to\snr_1\snr_1$.

The decay width of a scalar Higgs into a RH sneutrino pair  or a RH neutrino pair is \cite{Cerdeno:2011qv},
\begin{eqnarray}
	\Gamma_{\higgsi\to \snr_1\snr_1}&=&\frac{|\chsnsn |^2}{32\pi m_{H^0_i}}
	\left(1-\frac{4\snmassr^2}{m^2_{\higgs{i}}}\right)^{1/2}\,,\\
	\label{gammasnsn}
	\Gamma_{\higgsi\to\rhn\rhn}&=&\frac{\lambda_N^2(S^3_{\higgsi})^2}{32\pi}m_{\higgsi}\left(1-	\frac{4\rhnmass^2}{m^2_{\higgs{i}}}\right)^{3/2}\,,
	\label{gammann}
\end{eqnarray}
where the Higgs-sneutrino-sneutrino coupling reads \cite{Cerdeno:2009dv}
\begin{eqnarray}
	\chsnsn= \frac{2\lambda\lambda_N m_W}{\sqrt{2}g}\left(\sin{\beta}S^1_{\higgsi}+\cos{\beta}
	S^2_{\higgsi}\right)+\left[(4\lambda_N^2+2\kappa\lambda_N)v_s+
	\lambda_N\frac{A_{\lambda_N}}{\sqrt{2}}\right]S_{H^0_i}^3\,.
	\label{eq:c}
\end{eqnarray}
In terms of these, the branching ratio into invisible and non-SM channels reads,
\begin{eqnarray}
	\mbox{BR}{(\hsm\to inv)}= \frac{\Gamma_{\hsm\to inv}}{\Gamma_{NMSSM}+
	\Gamma_{\hsm\to inv}}\,,
\end{eqnarray}
where $\Gamma_{NMSSM}$ is the Higgs decay width in all other possible NMSSM products and is calculated using the code {\tt NMSSMTools}. $\Gamma_{\hsm\to inv}$ accounts for all non-standard decays of the Higgs boson, which in our model should comprise decays into pairs of RH neutrinos, RH sneutrinos, neutralinos, and scalar and pseudoscalar Higgs bosons, i.e., 
$\Gamma_{\hsm\to inv} = \Gamma_{\hsm\to \snr_1\snr_1} + \Gamma_{\hsm\to\rhn\rhn} + \Gamma_{\hsm\to\neut_i\neut_i} + \Gamma_{\hsm\to\higgsl\higgsl} + \Gamma_{\hsm\to\phiggsl\phiggsl}$.
In the scenarios considered in this work the neutralinos and lightest CP-even and CP-odd Higgses are heavier than $m_{\hsm}/2\approx62$~GeV, and therefore only the contributions from decays into RH neutrinos and sneutrinos are important. 

From the expressions above it is clear that if the decay into RH neutrinos is kinematically allowed then large values of $\ln$ can lead to a sizable contribution to the invisible decay, being therefore very constrained. On the other hand, regarding the Higgs decay into two RH sneutrinos, the Higgs-sneutrino-sneutrino coupling is a more complicated function, involving $\ln$, $\aln$, and $\mn$, and accidental cancellations might occur. In general, however, large $\ln$ is also more constrained.

We have constructed a chi-squared function, $\chi^2(\mu)$, for the total visible signal strength, $\mu$, using the data for the
signal strengths of each individual process
given by ATLAS and CMS. In order to be conservative we assume that $\mu=1-\mbox{BR}(h_{SM}^0\to inv)$, which holds if the Higgs is totally SM-like except for the new decays. This means that new contributions (apart from those of the SM) to the Higgs production are assumed to be zero. Although this is not always true for SUSY models, this implies a stronger bound on the invisible Higgs branching ratio. The minimum of the function is achieved for a non-zero value of the invisible Higgs branching ratio, and the 1$\sigma$ and 2$\sigma$ values are given by $\chi^2=\chi^2_{min}+\Delta \chi^2$, with $\Delta \chi^2=1$, and $4$ respectively. With this prescription, we obtain $\mbox{BR}{(h_{SM}^0\to inv)}<0.15(0.27)$ at $1\sigma(2\sigma)$, consistent with other recent analyses \cite{Espinosa:2012vu,Belanger:2013kya,Falkowski:2013dza,Giardino:2013bma,Ellis:2013lra,Djouadi:2013qya,Belanger:2013xza}.

\begin{figure}
        \includegraphics[scale=0.405]{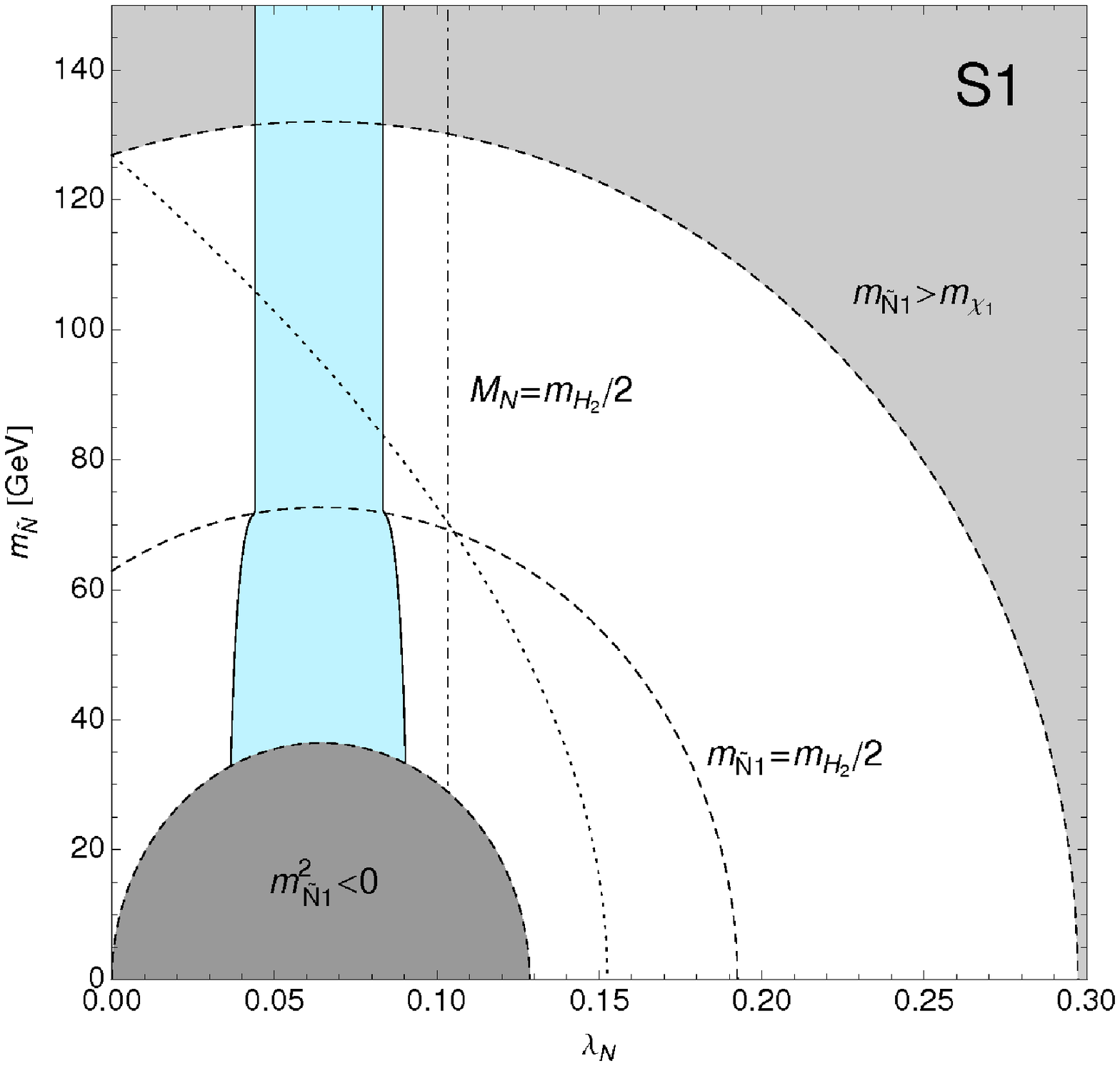}\hspace*{1cm}
        \includegraphics[scale=0.405]{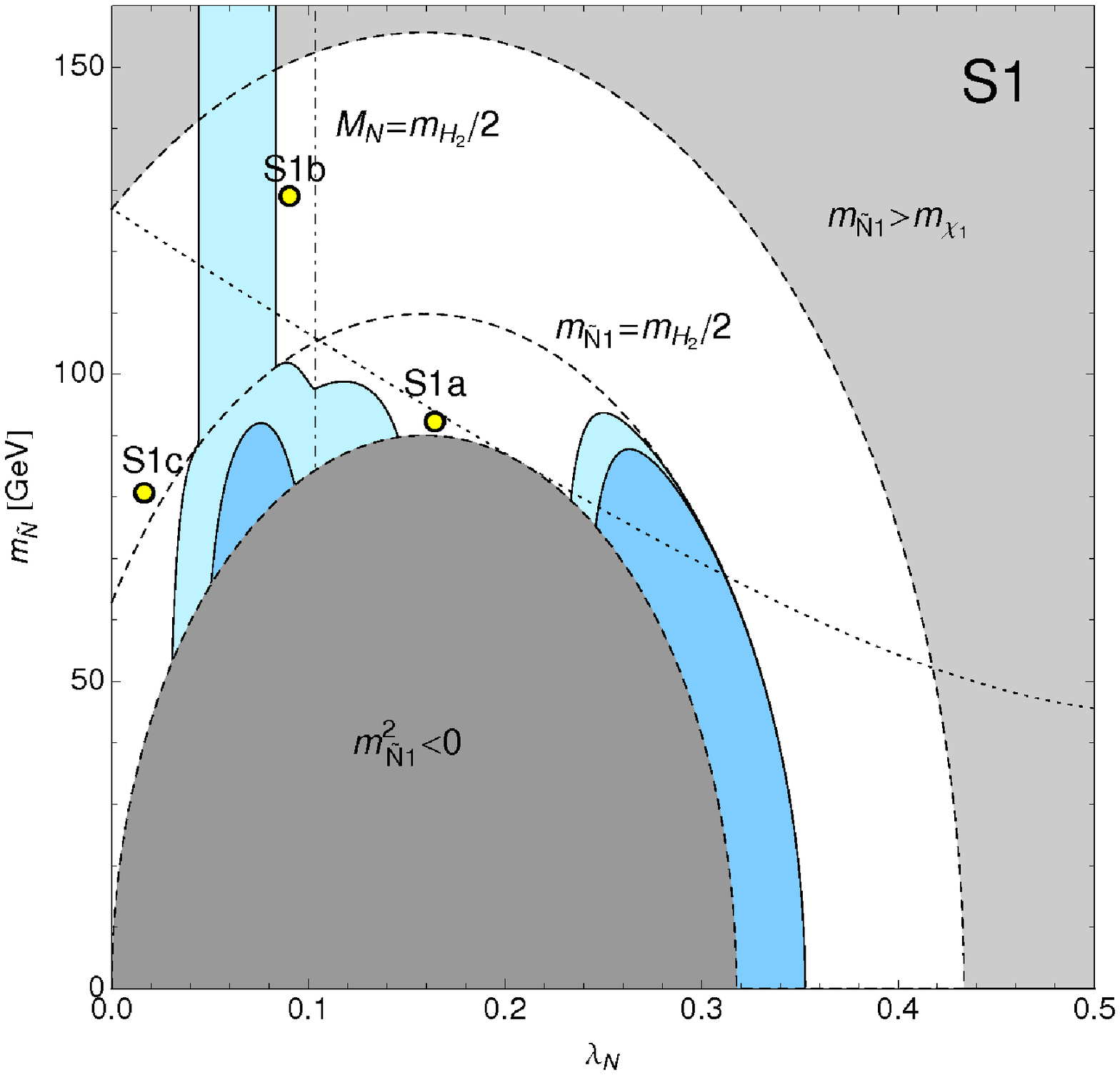}\hspace*{1cm}
\\[2ex]
        \includegraphics[scale=0.405]{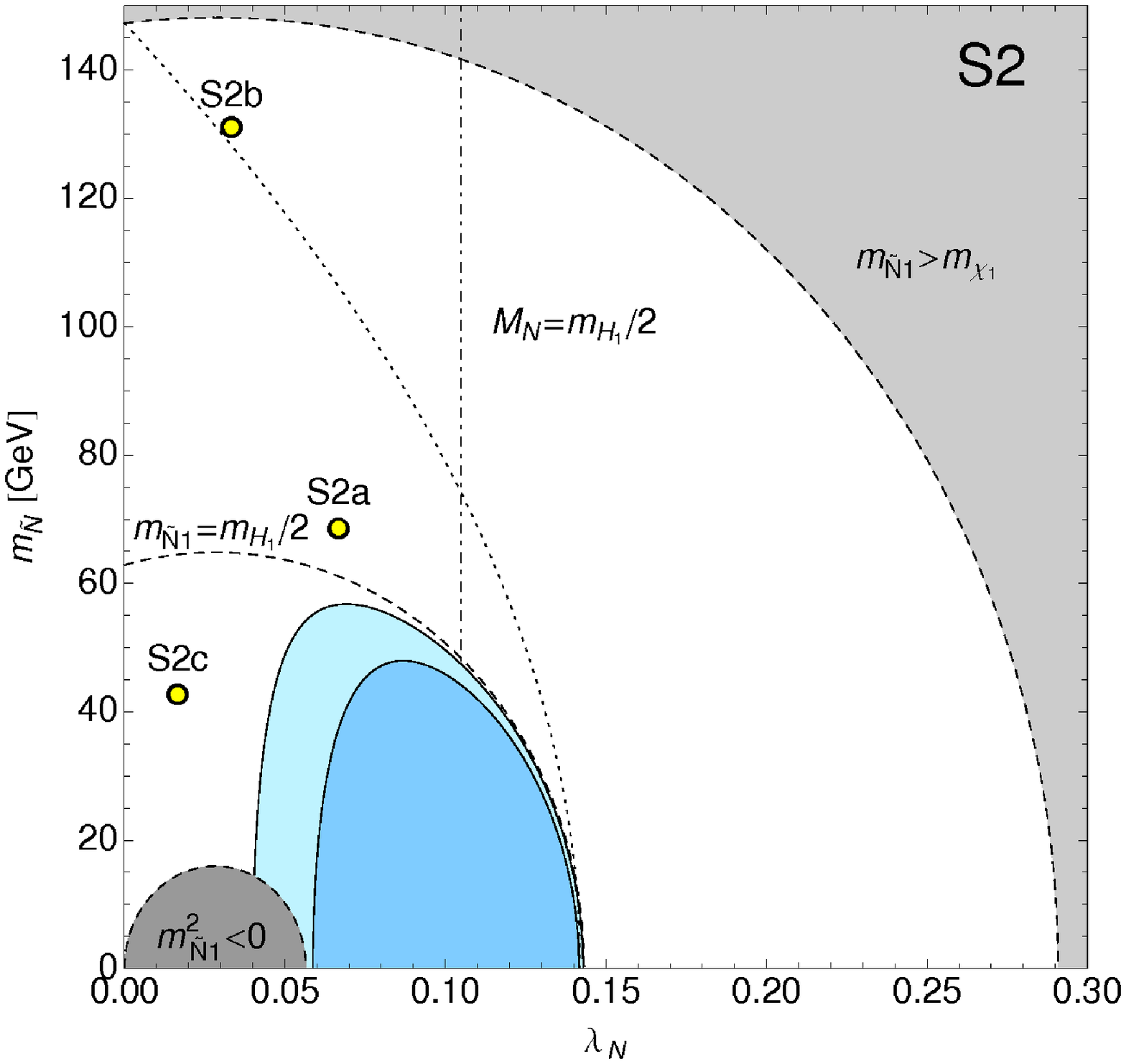}\hspace*{1cm}
        \includegraphics[scale=0.405]{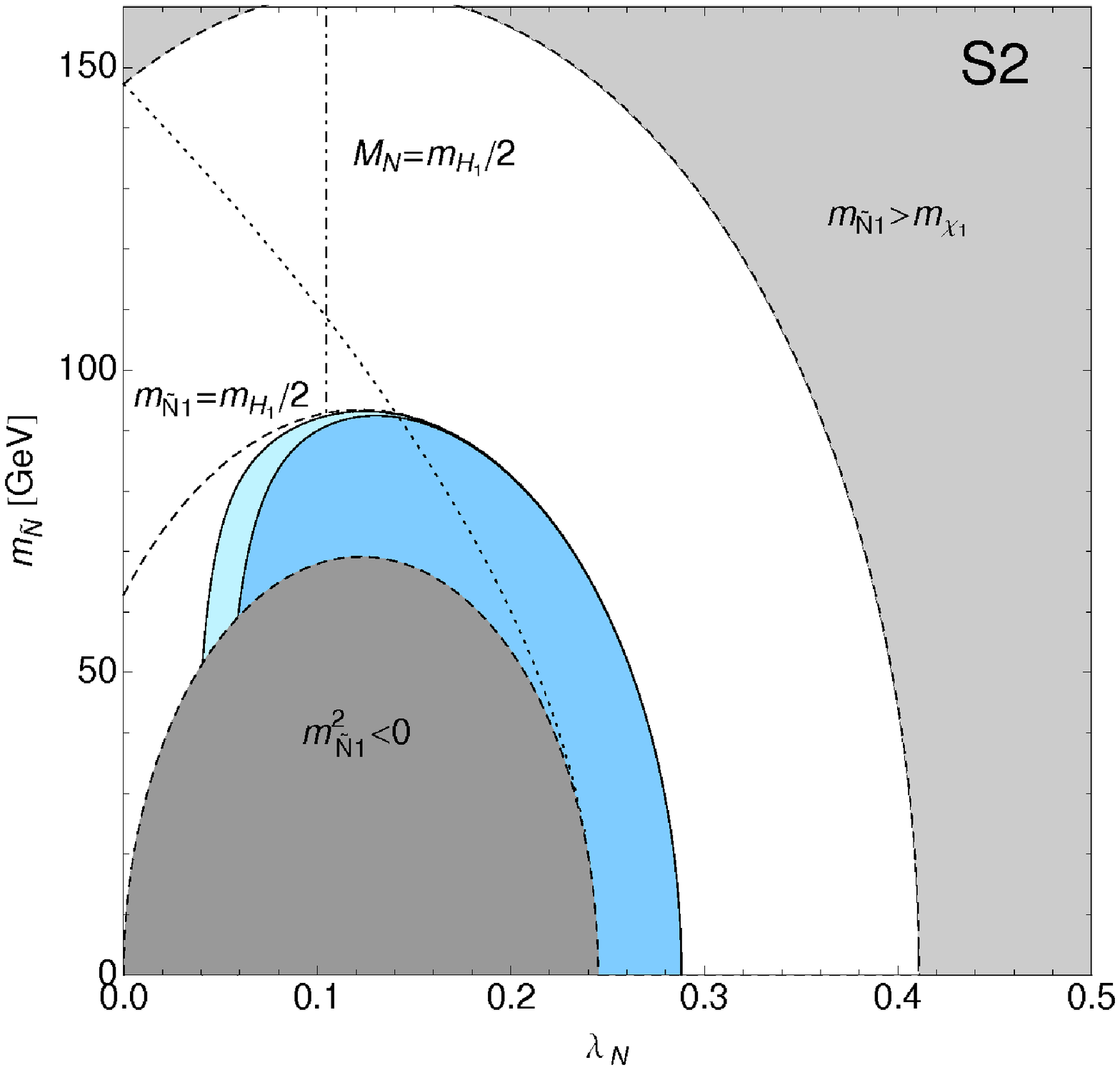}\hspace*{1cm}
        \caption{\small
Constraints on the $(\ln,\,\mn)$ plane from the invisible branching ratio of the SM-like Higgs for S1 (upper row) and S2 (lower row). From left to right, the  trilinear term is $\aln=-150$, $-250$~GeV. Dark (light) blue areas correspond to the regions of the parameter space where BR$({h_{SM}^0\to inv})>0.15 (0.27)$ , corresponding to the $1\sigma$ and $2\sigma$ exclusion limit by ATLAS and CMS. Dark gray areas are ruled out since the RH sneutrino mass-squared is negative. Dashed lines indicate the curves along which the RH sneutrino mass is constant and $\snmassr=\neutmass$, $\snmassr=\hmasssm/2$ from top to bottom. The vertical dot-dashed line corresponds to $\rhnmass=\hmasssm/2$. Finally, points to the left and below the dotted line satisfy $\snmassr+\rhnmass<\neutmass$. Yellow dots correspond to the various benchmark points, defined in Table\,\ref{tab:S-12}, that are used in the analysis.}
\label{fig:invisible}
\end{figure}

We illustrate in Fig.\,\ref{fig:invisible} the effect of these bounds on the $(\ln,\,\mn)$ plane corresponding to scenarios S1 and S2 of Table \ref{tab:S-12}. For each of these we consider two examples with  $\aln=-150$~GeV and $-250$~GeV. 
The light (dark) blue area corresponds to the region excluded due to an excess in the invisible branching ratio of the SM-like Higgs.
The constraints from the invisible Higgs decay are very dependent on the RH sneutrino parameters.
In Fig.\,\ref{fig:invisible} the RH sneutrino mass increases with the soft mass parameter $\mn$ (along semicircular trajectories that depend on $\ln$). The dark gray area corresponds to regions of the parameter space for which $\snmassrsq<0$ 
and the light gray area is the one with $\snmassr>\neutmass$, above which the RH sneutrino is no longer the LSP.
Dashed lines correspond to  trajectories with a constant $\snmassr$. In all the examples we observe that $\Gamma_{\hsm\to\snr_1\snr_1}$ becomes larger when $\aln$ and $\ln$ increase and more regions are excluded. 
Above the line with $\snmassr=\hmasssm/2$ the Higgs cannot decay into a RH sneutrino pair and is therefore less constrained.
On the other hand, the RH neutrino mass increases with $\ln$ and so does the decay width $\Gamma_{\hsm\to\rhn\rhn}$ (see eq.\,(\ref{gammann})). This gives rise to a vertical excluded area  for S1 in the range $0.04\lsim\ln\lsim0.09$ which is independent of $\aln$. In example S2  the decay width $\Gamma_{\hsm\to\rhn\rhn}$ is reduced since the SM-like Higgs has a smaller singlet component and therefore does not violate the experimental bound. The vertical dot-dashed line corresponds to $\rhnmass=\hmasssm/2$ so to the right of this line the Higgs decay into a RH neutrino pair is kinematically forbidden.

Finally, points in the area to the left and below the dotted line satisfy $\snmassr+\rhnmass<\neutmass$. In this area the neutralino NLSP can undergo the two-body decay $\neutl\to\snr_1\rhn$, whereas to the right and above the dotted line the dominant decay is $\neutl\to\snr_1 \nu_L$.

We have selected various representative benchmark points for each scenario, which are indicated in the plot by means of a yellow dot, labelled as S1a, S1b, S1c, S2a, S2b, and S2c, and with parameters defined at the bottom of Table \ref{tab:S-12}.

\section{Displaced vertices from late decaying RH neutrinos}
\label{sec:displaced}

\subsection{RH neutrino production}
\label{sec:production}
RH neutrinos can be produced at the end of a decay chain together with a RH sneutrino, when the latter is the LSP. 
If the wino-like neutralino and wino-like chargino are light, the leading production channel is $pp\to\neuti\charg_j$ (through a very off-shell $W$). Both neutralino and chargino subsequently decay into the RH sneutrino LSP in very short chains (e.g., $\neuti\to \snr_1\rhn$ and $\charg_j\to W^\pm\neut_1\to W^\pm\snr_1\rhn$).

RH neutrinos can also be produced directly in the decay of a scalar Higgs boson. This is a very clean channel, however it can be suppressed. On the one hand, the production of a Higgs particle is proportional to its doublet component (which determines the Higgs coupling to SM particles), but the decay of the Higgs into RH neutrinos is only sensitive to its singlet component. Notice also that the $\hsm\to\rhn\rhn$ branching ratio is also constrained to be small from the recent bounds on invisible Higgs decays.

\begin{figure}
        \includegraphics[scale=0.73]{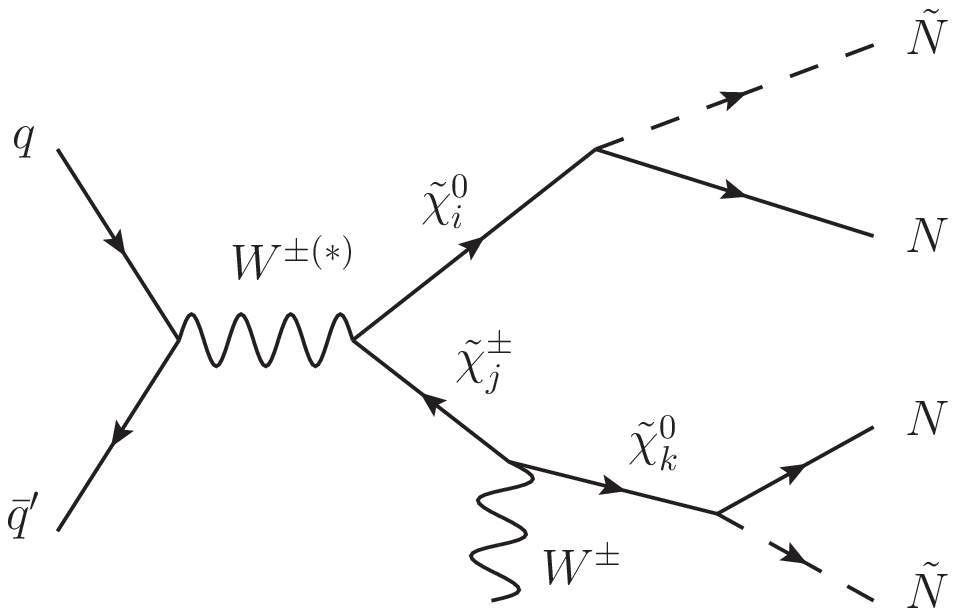}\hspace*{4ex}
        \includegraphics[scale=0.75]{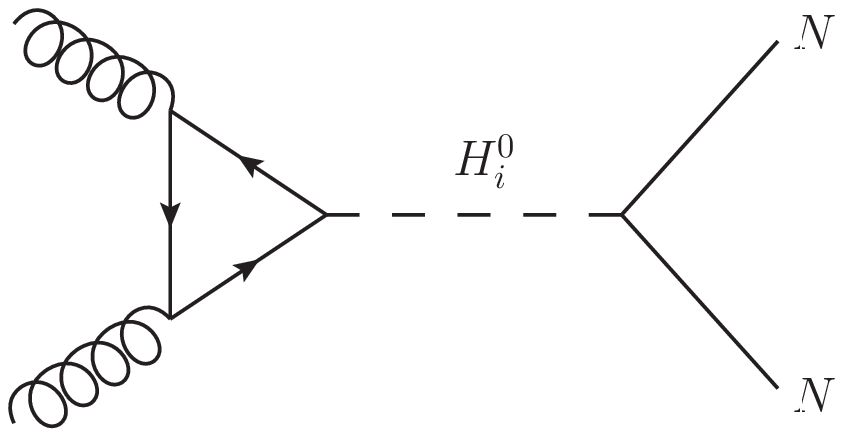}
	\caption{\small Different possibilities for the production of RH sneutrinos. On the left, a neutralino/chargino pair is produced after the original collision and undergoes a short decay chain that ends in the production of a RH neutrino/sneutrino. On the right, a pair of  RH neutrinos is produced in the decay of a Higgs boson.}
	\label{fig:rhn-production}
\end{figure}

Both production mechanisms are illustrated in Fig.\,\ref{fig:rhn-production}.  The RH neutrino eventually decays into Standard Model particles. Notice that depending on the masses of the particles involved, the RH neutrino can be produced on-shell or be an off-shell mediator of higher order decays. We can define three potential scenarios.

\begin{itemize}
\item[(I)] $\rhnmass<\neutmass-\snmassr$

In this case, the lightest neutralino two-body decay $\neutl\to\snr_1\rhn$ is kinematically allowed. This proceeds very rapidly, since the coupling $\csnnneui$ is not Yukawa suppressed. In this case, the RH neutrino and sneutrino are produced on-shell and 
the on-shell RH neutrino can be long-lived. 

\item[(II)] $\neutmass-\snmassr<\rhnmass<\neutmass+\snmassr$

The neutralino two-body decay $\neutl\to\snr_1\rhn$ is not possible, but it can proceed through a virtual RH neutrino into a multi-body final state, where the virtual RH neutrino vertex introduces a factor $\yn^2$ in the total neutralino decay width. 
However, the two-body decay $\neutl\to\snr_1\nu_L$ is always kinematically allowed and dominates the decay width. Although it is suppressed by the mixing between the left and right neutrino components (and therefore also introduces a factor $y_N^2$), it is favoured by the phase space with respect to the possibility discussed above.
Since the decay products of the neutralino are invisible, this scenario does not leave any displaced vertex (and is indistinguishable from the production of neutralino dark matter). 
This implies that for this range of RH neutrino masses we cannot consider the production mechanism through a neutralino-chargino pair.

On the other hand, this does not affect RH neutrinos produced through Higgs decays. 

\item[(III)] $\neutmass+\snmassr<\rhnmass$

Finally, if RH neutrinos are heavy enough that the decay channel $\rhn\to\neutl\snr_1$ is kinematically allowed, then no displaced vertices are expected, since $\snr_1$ is stable and, as explained above, $\neutl\to\snr_1\nu_L$ is the dominant decay channel for the lightest neutralino.

\end{itemize}

\subsection{RH neutrino decays}
\label{sec:rhn}

\begin{figure}
        \includegraphics[scale=0.39]{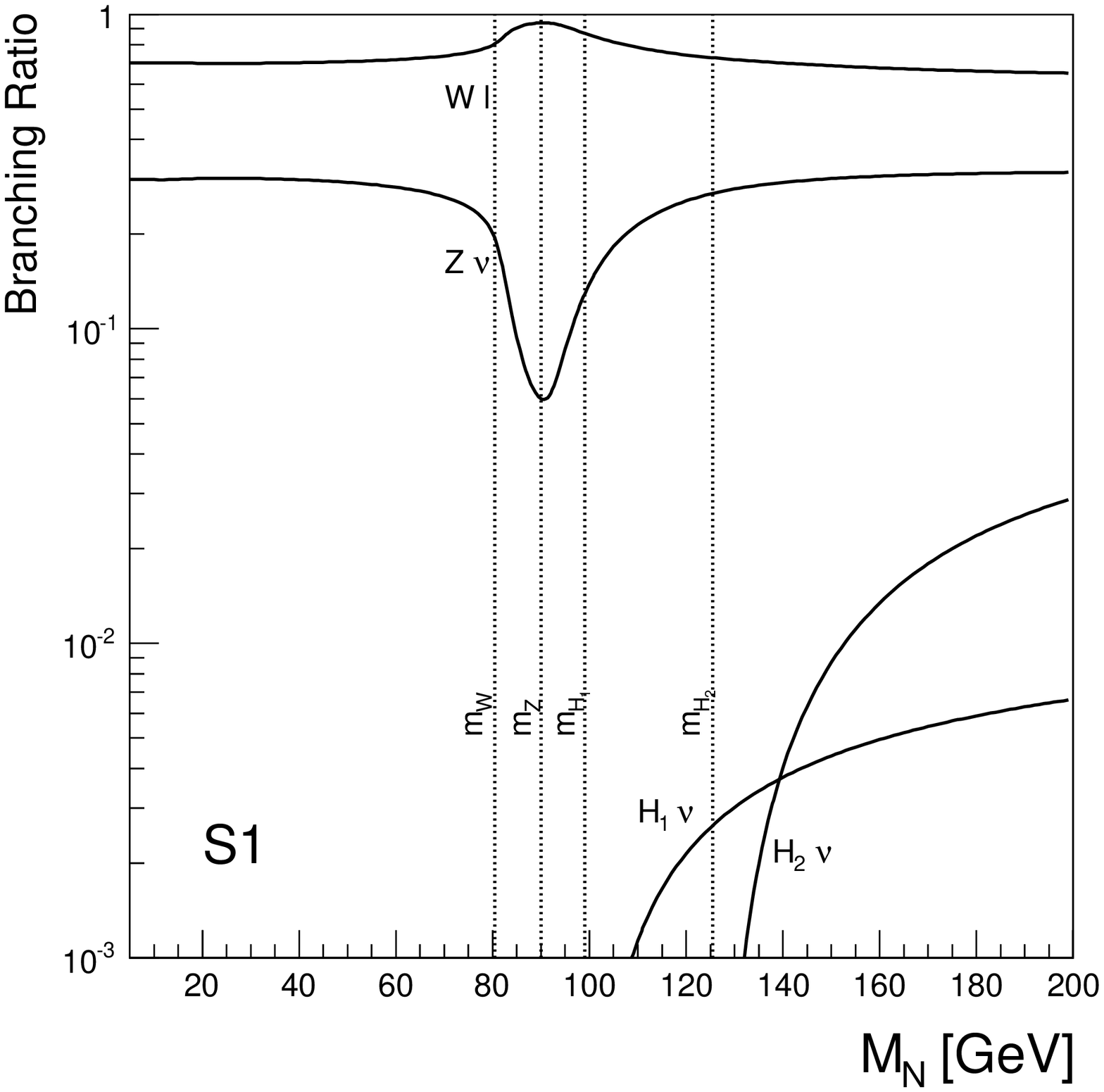}\hspace*{1.ex}
        \includegraphics[scale=0.39]{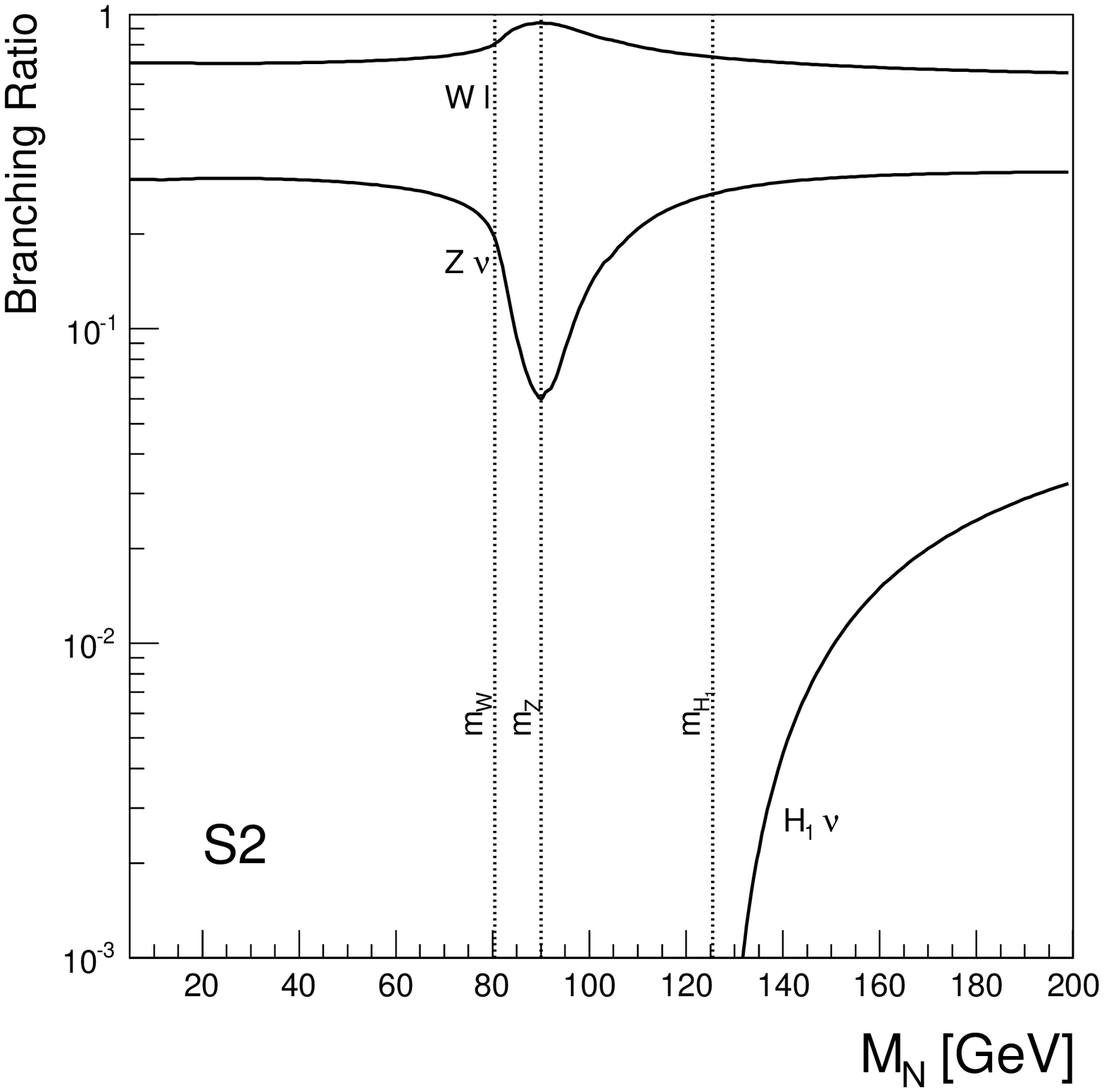}
	\caption{\small Branching ratios of the RH neutrino as a function of its mass for scenario S1 (left) and S2 (right). }
	\label{fig:rhn-BR}
\end{figure}
\begin{figure}[!t]
        \includegraphics[scale=0.39]{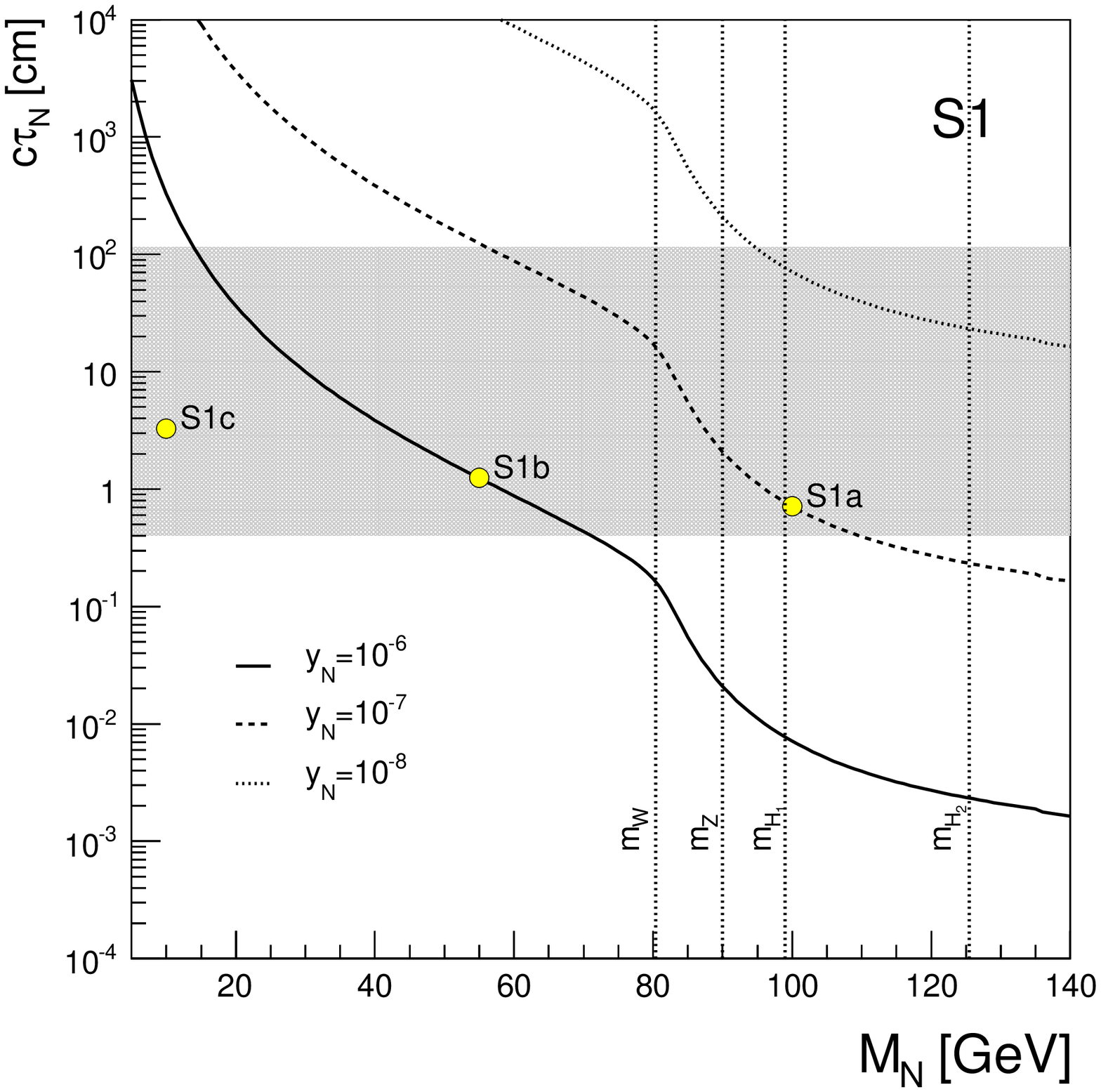}\hspace*{1.ex}
        \includegraphics[scale=0.39]{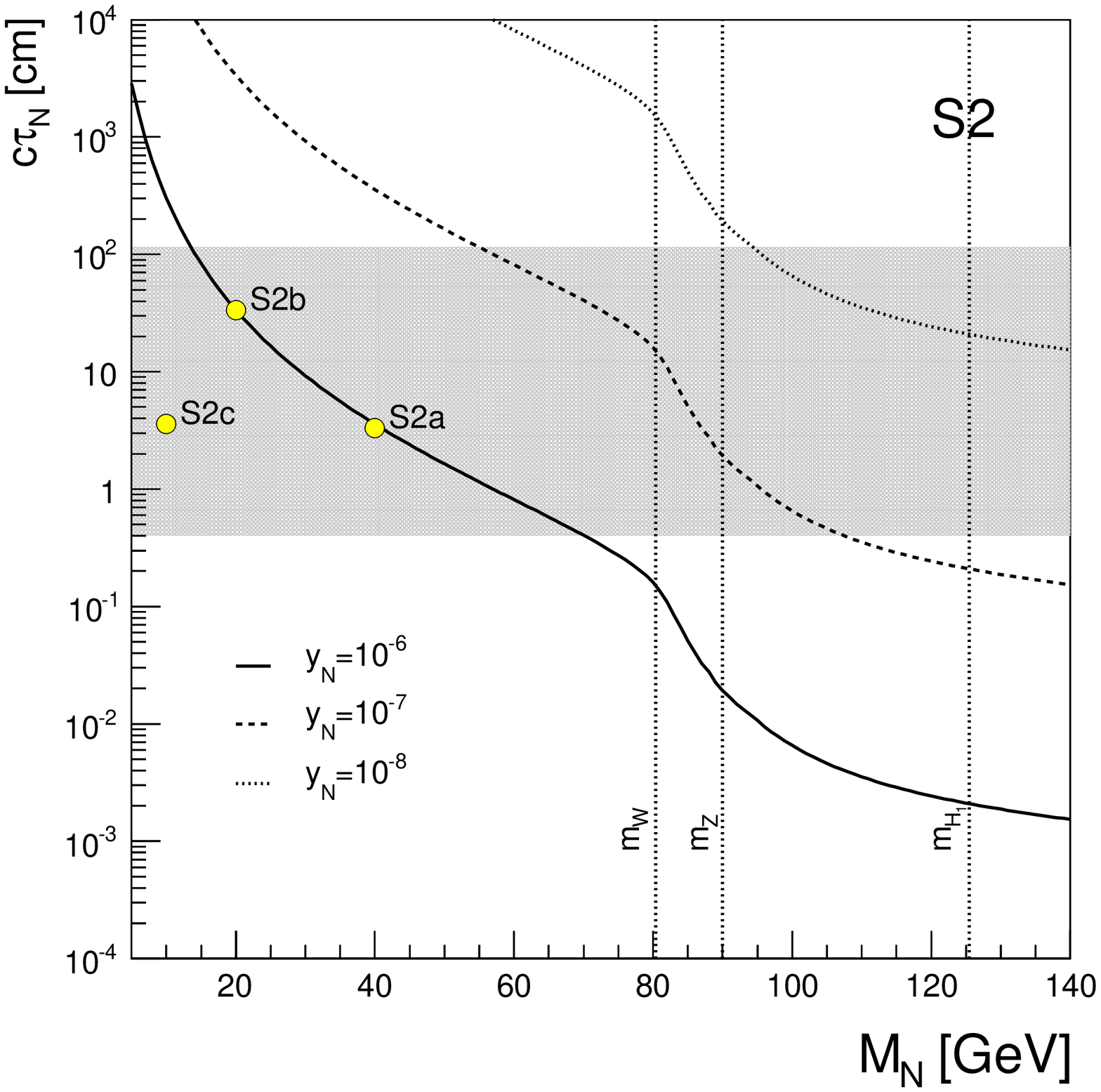}
        \caption{\small Decay length of the RH neutrino as a function of its mass for benchmark points S1 (left) and S2 (right). It is important to note that the decay length is independent of the sneutrino mass. The different lines represent different values of  the neutrino Yukawa coupling. The shaded area corresponds to the range in lengths that could lead to an observable displaced vertex in the ATLAS inner detector.}
         \label{fig:ctau}
\end{figure}

If the RH neutrino is heavy enough, it can undergo a two-body decay into $W^\pm l^\mp$, $Z\nu_L$, or $\higgsi\nu_L$. The decay width corresponding to all these channels is proportional to $\yn^2$, which enters either through the LR mixing of the neutrino (in the cases $N\rightarrow W^\pm l^\mp$ and $N\rightarrow Z\nu_L$) or in the coupling with the Higgs (in the case of $N\rightarrow \higgsi\nu_L$), 
\begin{eqnarray}
	\Gamma_{\rhn\to W l}&=& \frac{y_N^2 v_2^2 g^2}{64\pi}\frac{\rhnmass}{m_W^2}
	\left(1-\frac{m_W^2}{\rhnmass^2}\right)^2\left(1+\frac{2m_W^2}{\rhnmass^2}\right)\,,\\
	\Gamma_{\rhn\to Z \nu_L}&=& \frac{y_N^2 v_2^2 g^2}{64 \pi}\frac{\rhnmass}{m_W^2}
	\left(1-\frac{m_Z^2}{\rhnmass^2}\right)^2\left(1+\frac{2m_Z^2}{\rhnmass^2}\right)\,,\\
	\Gamma_{\rhn\to \higgsi\nu_L}&=& \frac{y_N^2(S^2_{\higgsi})^2}{8\pi}\rhnmass
	\left(1-\frac{m_{\higgsi}^2}{\rhnmass^2}\right)^2\,,
\end{eqnarray}
where $S^2_{\higgsi}$ is the up component of the Higgs $\higgsi$.
Therefore, we expect this particle to be long-lived, and give rise to a displaced vertex that could be observed through the resulting charged SM particles. Notice in this sense that the channels $Z\nu_L$ and $\higgsi\nu_L$ are only observable through the decay products of the $Z$ and $\higgsi$ bosons.

For lighter $\rhn$, we can only have three-body decays through virtual $W^\pm$, $Z$ or $\higgsi$. For the same reasons as above, the decay width is proportional to $\yn^2$ but now is further suppressed by the phase space, thus leading to a larger lifetime. We have computed the corresponding lifetime using {\tt CalcHEP 3.4}.

Thus, in terms of the parameters of the model, the RH neutrino lifetime is only a function of its mass, sensitive to the details of the Higgs sector, and modulated by $\yn^2$. We illustrate the results with two numerical examples, denoted benchmark points S1 and S2, with parameters defined in Table\,\ref{tab:S-12}. 
The resulting decay length and branching ratios are displayed as a function of the RH neutrino mass on the left and right panels of Fig.\,\ref{fig:rhn-BR} and Fig.\,\ref{fig:ctau}, respectively. The shaded area corresponds to the range in distances that we expect the ATLAS inner detector can resolve for a displaced vertex. 
We indicate by means of dotted vertical lines the masses of the gauge bosons and $\higgsl$, below which two-body decays are no longer possible.

As we observe, for a wide range of values for the RH neutrino mass and the neutrino Yukawa,  the RH neutrino decay length is within the range that can be resolved in ATLAS. 
Also, we have found that in general the decay through a virtual or on-shell $W$ dominates the decay width, and this will determine our choice of signals to analyse.

\subsection{Signals at the LHC}
\label{sec:summary-displaced}

\begin{table}
\begin{center}
\begin{tabular}{|c|c|c|c|}
\hline
Process & Signature\\
\hline
$N\rightarrow  W^\pm l^{\mp}_i\rightarrow \nu_j l^\pm_j l^{\mp}_i  $ & $2\ell (+ \met)$
\\
$\phantom{N\rightarrow  W^\pm l^{\mp}_i}\rightarrow  q \bar{q}'  l^{\mp}_i$ & $\ell jj$
\\
$N\rightarrow  Z \nu_i\rightarrow \nu_i l^\pm_j l^{\mp}_j  $ & $2\ell (+ \met)$
\\
$\phantom{N\rightarrow  Z \nu_i}\rightarrow \nu_i q \bar{q}  $ & $2 j (+ \met)$
\\
$N\rightarrow  H^0_i \nu_i\rightarrow \nu_i l^\pm_j l^{\mp}_j  $ & $2\ell (+ \met)$
\\
$\phantom{N\rightarrow  H^0_i \nu_i}\rightarrow \nu_i q \bar{q}  $ & $2 j (+ \met)$
\\
$\phantom{N\rightarrow  H^0_i \nu_i}\rightarrow \nu_i \gamma \gamma  $ & $2 \gamma (+ \met)$
\\
\hline
\end{tabular}
\end{center}
\caption{\small Potential signatures at the LHC corresponding to the different production mechanisms, where $\ell = e^\pm,\,\mu^\pm$ and $j$ stands for hadronic jets. As explained in the text, it is difficult to measure the missing transverse energy, $\met$, associated with a displaced vertex.}
      \label{tab:signals}
\end{table}

The displaced vertex originating from the late decay of a RH neutrino is observable through the decay products of the $W^\pm$, $Z$, and $\higgsi$ bosons.

The observation of a displaced vertex depends on the reconstruction of the tracks of the charged particles produced. Usually at least two charged tracks are needed to reconstruct a secondary vertex. Important parameters for the reconstruction are the total distance from the primary vertex, $L_{xy}$, and the impact parameter, $d_0=L_{xy}\sin \theta$, where $\theta$ is the angle described by the trajectory of the displaced vertex with respect to the beam line. The resolution of the vertices in the pixel tracker for both ATLAS and CMS detectors is of the order of hundred $\mu$m, however as the displaced vertices must be distinguished from primary ones, usually $|d_0|\gtrsim 2-4$ mm and $L_{xy}\gtrsim 4-8$ mm is imposed \cite{ATLAS:2012av, Aad:2012kw, Aad:2012zx, Chatrchyan:2012jna}. These cuts also remove completely the SM background, as it was shown in Refs. \cite{Aad:2012kw,Aad:2012zx,Chatrchyan:2012jna,Bobrovskyi:2012dc}. 
In fact, in Ref.\,\cite{Chatrchyan:2012jna} it was shown through MC simulation studies that the selection on $|d_0| > 2$mm can remove 98\% of all tracks from the primary $pp$ vertices.
Although in their searches for displaced vertices ATLAS and CMS use the whole detector volume, the identification of a displaced vertex decreases when $d_0$ grows \cite{ATLAS:2012av, Aad:2012kw, Chatrchyan:2012jna}. In our analysis we will therefore restrict our searches to the inner detector\footnote{For CMS(ATLAS) the inner detector has a radius of 110(108.2)\,cm  \cite{Bayatian:2006zz, Aad:2009wy}.} and for each simulated event we impose impose a cut on the decay length of the RH neutrino, 10 mm$<c\tau_\rhn<$100 cm.

The results for current searches for displaced vertices using the ATLAS or CMS detector can be found in Refs. \cite{ATLAS:2012av, Aad:2012kw, Aad:2012zx} and Ref. \cite{Chatrchyan:2012jna}, respectively. The efficiencies shown there are dependent on the distance where the displaced vertex takes place. In some points the reconstruction efficiency could be as large as 30\%, but in general this efficiency is smaller.

In Table\,\ref{tab:signals} we detail the potential signatures\footnote{Similar signatures have been described in singlino decays in the $\mu\nu$SSMÊ\cite{Bartl:2009an}.}
As mentioned in the previous section, we expect the contribution from $\rhn\to W^\pm l^\mp$ to be dominant and therefore we concentrate on the two signatures $\rhn\to 2\ell +\met$ and $\rhn \to \ell jj$.
Notice however that $\rhn\to 2\ell +\met$ also receives contributions from processes in which the mediator is either the Higgs or the $Z$ boson and this will be taken into account. 
It is important to observe that the emitted neutrinos contribute to the missing energy of the total event. The missing energy cannot be associated to the displaced vertex itself (as the neutrino cannot be detected). For this reason, $\met$ is not a good variable in our study and we only focus on the properties of the visible particles that originate in the displaced vertices (leptons and jets).

\begin{itemize}

\item $2l(+\met)$

In principle one could think of using the transverse mass, defined as
\begin{equation}
m_{T}^2=\left(\sqrt{M^2_{vis}+\vec{p}_{{T},vis}^{\,2}} + \met^V\right)^2-\left(\vec{p}_{{T},vis}+\vec{\slashed{p}}_{T}^V\right)^2,
\end{equation}
where $M^2_{vis}$ is the invariant mass of the visible system, $\vec{p}_{{T},vis}$ is the transverse momentum vector of the visible system, $\met^V$ is the missing transverse energy of the vertex and $\vec{\slashed{p}}_{T}^V$ is the vector of the missing transverse energy. However, in practice we would not be able to measure the missing transverse energy that comes from the displaced vertex. Notice for example that it would be very difficult, if not impossible, to determine the original interaction from which the long-lived RH neutrinos originated.

For this reason, we try to make use of only the visible particles. It was shown in Refs.\cite{Paige:1996nx, Nojiri:1999ki, Gjelsten:2004ki} that the invariant mass for the dileptonic system presents an endpoint which is sensitive to relations between the particles involved. When applied to the process $N\to Wl/Z\nu_L/\higgsi\nu_L\to ll\nu_L$, it can be shown that if the intermediate particle is produced on-shell, an edge in the resulting distribution will be present for, 
\begin{equation}
\left(m^{edge}_{l_1 l_2}\right)^2=(\rhnmass^2-m_X^2)\, ,
\end{equation}
where $X=W,Z,\higgsi$. If the intermediate particle is produced off-shell, the distribution is expected to have an endpoint at
\begin{equation}
\left(m^{end-point}_{l_1 l_2}\right)^2=(\rhnmass-m_{\nu_L})^2\simeq \rhnmass^2.
\end{equation}

Since there are different intermediate particles for the decay of the RH neutrino, one expects various edges, which might be difficult to distinguish. Also, the invariant mass of two leptons will have resonant peaks for the $Z$ and for the $\higgsi$. 
We can avoid these two problems if we only consider the final states coming from the $W$ boson. This can be done by requiring two leptons with different flavours that arise from the same displaced vertex. We thus eliminate leptons coming from the $Z$ and $\higgsi$ channels that could spoil the mass reconstruction. Furthermore, as mentioned above, the  $W$ boson decay is favoured being the main branching ratio of the RH neutrino.

\item $ljj$

A plausible strategy to obtain information about the RH neutrino that produces the displaced vertex is to analyse the two jets plus the lepton arising from the same vertex. If we are able to reconstruct these three elements it is easy to calculate the invariant mass of the system, defined as
\begin{equation}
	m^2_{jjl}=(p^{\mu}_{j_1}+p^{\mu}_{j_2}+p^{\mu}_{l})(p_{j_1\mu}+p_{j_2\mu}+p_{l\mu}),
\end{equation}
where $p_i^\mu$ are the Lorentz vectors of the different particles.  Since the decay width of the RH neutrino is much smaller than its mass $\Gamma_{N}\ll \rhnmass$, this variable presents a kinematical peak near the pole mass of the RH neutrino.

\end{itemize}

\begin{table}[!t]
\begin{center}
\begin{tabular}{|l||c|c||c|c||c|c|c|}
\hline
&$\sigma^{8\text{TeV}}_{\higgsi}$ &$\sigma^{8\text{TeV}}_{\tilde{\chi}_j^\pm\tilde{\chi}_i^0}$ &$\sigma^{13\text{TeV}}_{\higgsi}$ &$\sigma^{13\text{TeV}}_{\tilde{\chi}_j^\pm\tilde{\chi}_i^0}$ &$\rhnmass$ &$\snmassr$&$m_{\tilde{\chi}_1^0}$ \\
\hline
S1a&2$\times 10^{-5}$&0.87&3$\times 10^{-4}$&1.94&100&20&127\\
S1b&0.89&--&2.06&--&55&100&127\\
S1c&0.54&0.87&1.24&1.94&10&70&127\\
\hline
S2a&0.004&0.25&0.22&0.65&40&70&147\\
S2b&0.034&--&0.48&--&20&130&147\\
S2c&0.009&0.25&0.29&0.65&10&40&147\\
\hline
\end{tabular}
\end{center}
\caption{\small Contributions to the production cross section of a RH neutrino pair from Higgs decays ($\sigma^{8\text{TeV}}_{\higgsi}$) and neutralino/chargino pair-production ($\sigma^{8\text{TeV}}_{\tilde{\chi}_j^\pm\tilde{\chi}_i^0}$ ) at the LHC with a centre of mass energy of 8 TeV and 13 TeV. For convenience, the masses of the particles involved are also indicated. The production cross sections are given in pb while the masses are in GeV. In benchmark points S1b and S2b the neutralino decay into RH neutrino and RH sneutrino is kinematically forbidden and neutralino/chargino production is not considered.}
      \label{tab:productionsigma}
\end{table}

We are not aware of algorithms that simulate the reconstruction of displaced vertices at the detector level. Thus we have carried out our data simulation at parton level using {\tt CalcHEP}. In order to simulate the detector effects on the reconstruction of the energies of leptons and jets, we assume that the nominal energies are smeared with a Gaussian distribution such that 
\begin{equation}
\frac{\sigma}{E}=\frac{a}{\sqrt{E/{\rm GeV}}}\oplus b\, ,
\label{eq:smearing}
\end{equation}
where $\oplus$ denotes sum in quadrature. For electrons we have $a_{\ell}=5\%$, and $b_{\ell}=0.55\%$, whereas jets are much more difficult to reconstruct and we will take $a_{j}=100\%$ and $b_{j}=5\%$ \cite{Aad:2009wy}. 
Muons are measured in the muon chambers and the smearing is applied to their transverse momentum.

For each of the benchmark points in scenarios S1 and S2 in Table \ref{tab:S-12} we have simulated the production of RH neutrinos in proton-proton collisions with the corresponding LHC configuration, considering the two production mechanisms detailed in Subsec.\,\ref{sec:production}. The production cross sections are specified in Table \ref{tab:productionsigma}.
Then, the generated event samples have been scaled to the given luminosity. We consider the current LHC values, with a centre of mass energy of 8~TeV and an integrated luminosity of $\mathcal{L}=20$ fb$^{-1}$, as well as a future scenario with 13~TeV and $\mathcal{L}=100$ fb$^{-1}$.

The following basic cuts are imposed in order to single out the signals.

\begin{itemize}
\item[-] In order to clearly discriminate the displaced vertices from $b$-jets, that usually have a $c\tau\sim 4$mm \cite{Bhattacharya:2011pc}, we require the displacement to be sufficiently large (but still contained within the inner detector). We thus require the presence of {\it two} displaced vertices with $10\,{\rm mm}<c\tau<100\,{\rm cm}$.
\item[-] For isolated electrons we require $p_T>10$~GeV and for muons $p_T>6$~GeV and $|\eta_{\ell}|<2.5$.
\item[-] For each jet we require $p_T>15$~GeV and $|\eta_{j}|<2.5$.
\item[-] The criterion for considering a particle or a jet isolated is $\Delta R>0.4$, where $\Delta R\equiv \sqrt{(\Delta \eta)^2 + (\Delta \phi)^2}$, with $\Delta \phi$ and $\Delta \eta$ being the azimuthal angular separation and the rapidity difference between two particles. We also make sure that the particles from one displaced vertex are isolated with respect to those of the other.
\end{itemize}

\begin{table}[!t]
\begin{center}
\begin{tabular}{|l|c|c|c|c|c||c|c|c|c|c|}
\hline
&\multicolumn{5}{|c||}{$\sqrt{s}=8$ TeV, $\mathcal{L}=20\, \text{fb}^{-1}$}&\multicolumn{5}{|c|}{$\sqrt{s}=13$ TeV, $\mathcal{L}=100\, \text{fb}^{-1}$}\\
\hline
&$ee$&$\mu\mu$&$e\mu$&$ejj$&$\mu jj$&$ee$&$\mu\mu$&$e\mu$&$ejj$&$\mu jj$\\
\hline
S1a&9&10&17&36&40&95&101&195&393&427\\
S1b&26&25&46&24&33&241&223&434&224&293\\
S1c&25&43&64&0&0&317&547&813&2&3\\
\hline
S2a&30&25&49&46&52&528&438&882&804&893\\
S2b&2&2&4&0&1&32&31&57&5&7\\
S2c&1&2&3&0&0&21&33&51&0&0\\
\hline
\end{tabular}
\end{center}
\caption{\small Number of events that pass all the cuts for the LHC configurations $\sqrt{s}$ = 8 TeV, $\mathcal{L}=20\, \text{fb}^{-1}$ and $\sqrt{s}$ = 13 TeV, $\mathcal{L}=100\, \text{fb}^{-1}$. An efficiency of 20\% is assumed in the reconstruction of displaced vertices.}
      \label{tab:events}
\end{table}

These cuts are designed in order to remove the SM model background. As it is shown in Refs. \cite{Aad:2012kw,Aad:2012zx,Chatrchyan:2012jna,Bobrovskyi:2012dc} the main SM background is due to $\gamma^* / Z^*\to \ell^+\ell^-, 
Z^*Z^*$. The cut imposed in the decay length is very effective and it can be seen that when it is combined with the condition that the invariant mass of two leptons are greater than 5 GeV, the SM background can be totally removed.  Our cut in the decay lenght is more restrictive so we make sure that we remove the SM background. We do not impose the cut on the invariant mass of the two leptons since in our scenarios the neutrinos are heavier than 5 GeV and a possible residual of background does not affect to the endpoint of the invariant mass distribution.

As it was pointed out before, the reconstruction efficiency of the displaced vertices is very poor. In our analysis we use the estimations for ATLAS and CMS and will assume that the efficiency is 20\%.

The number of signal events after all the cuts are applied is given in Table \ref{tab:events} for each benchmark point and each signal ($\ell\ell$ and $\ell j j $). We would like to remind the reader at this point that we are considering that the RH neutrino has equal mixings with the three left-handed neutrinos. Deviations from this assumption would imply variations in the relative rates for electron and muon signals.

\subsection{Results}

\begin{figure}[!t]
\hspace*{-4ex}
        \includegraphics[scale=0.28]{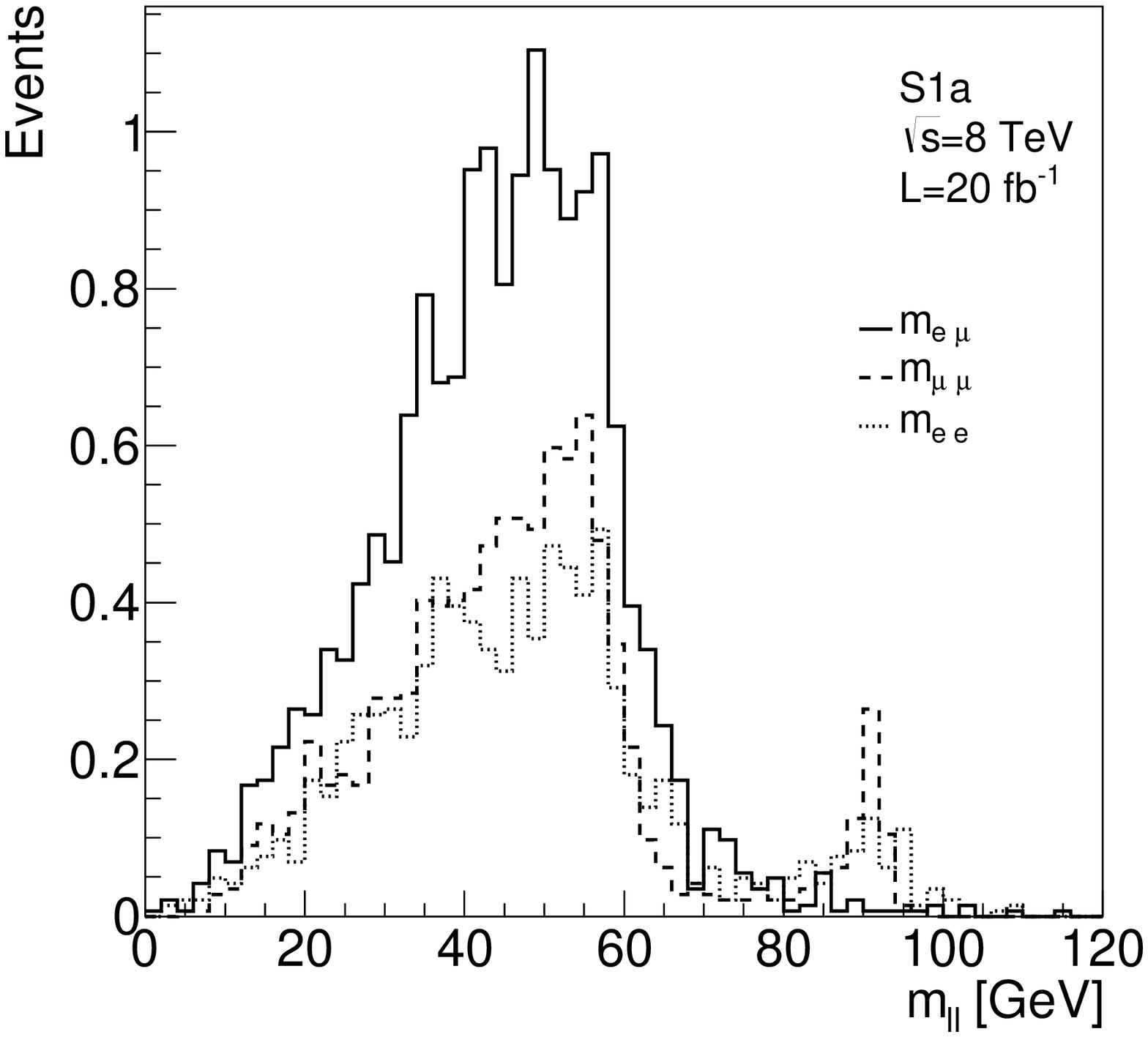}\hspace*{-1ex}
        \includegraphics[scale=0.28]{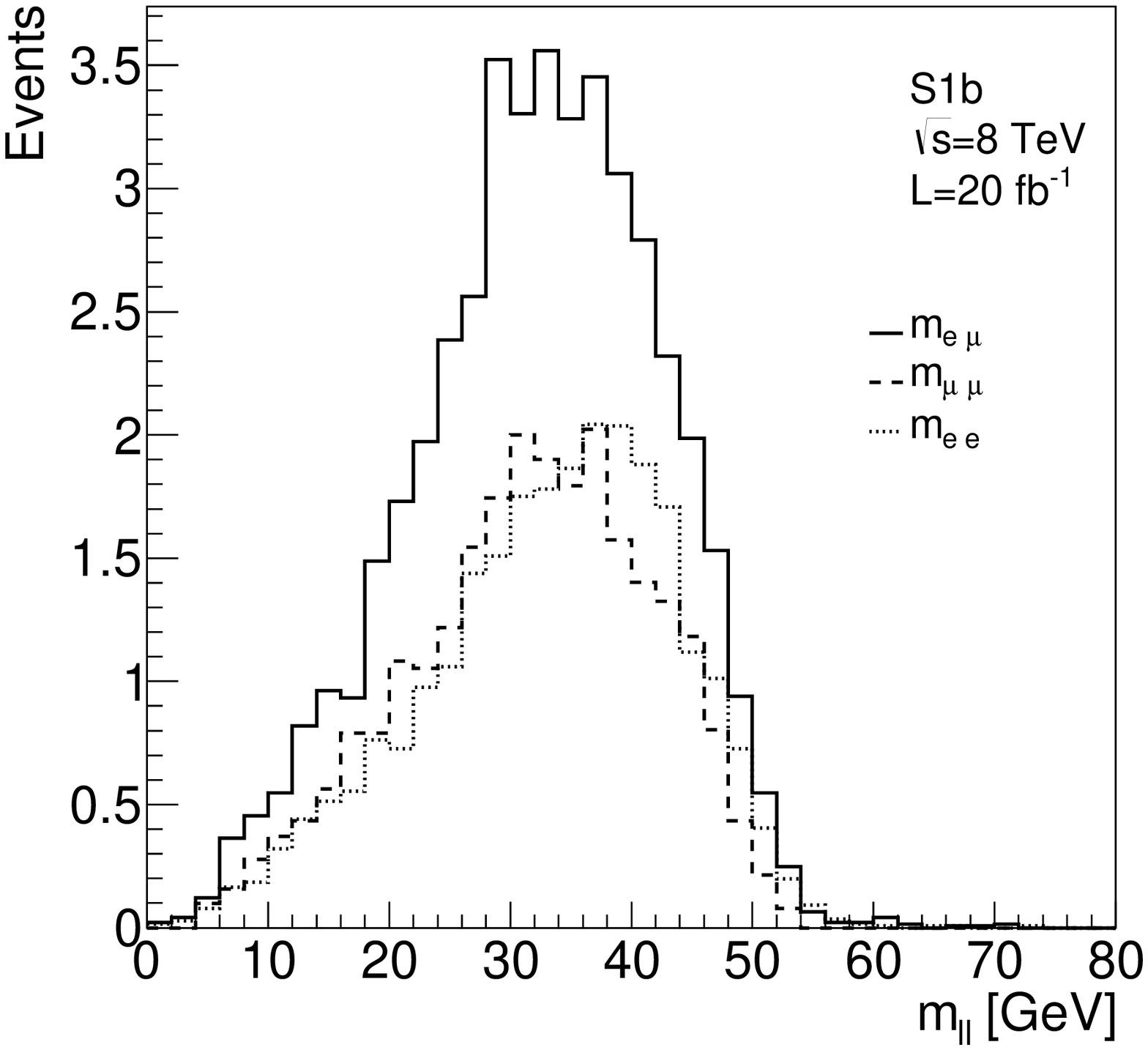}\hspace*{-1ex}
        \includegraphics[scale=0.28]{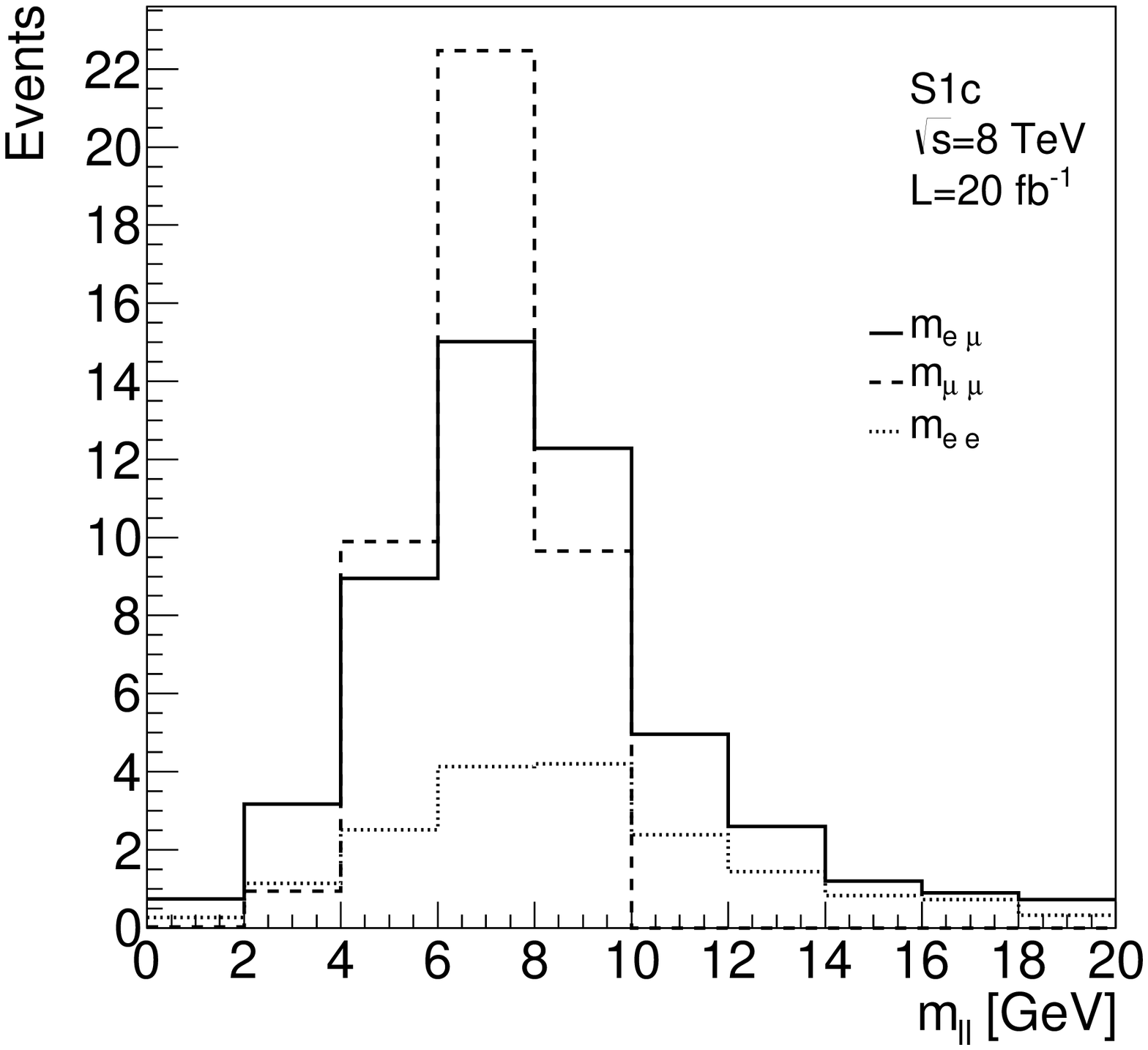}\hspace*{-1ex}\\      
\hspace*{-4ex}  
        \includegraphics[scale=0.28]{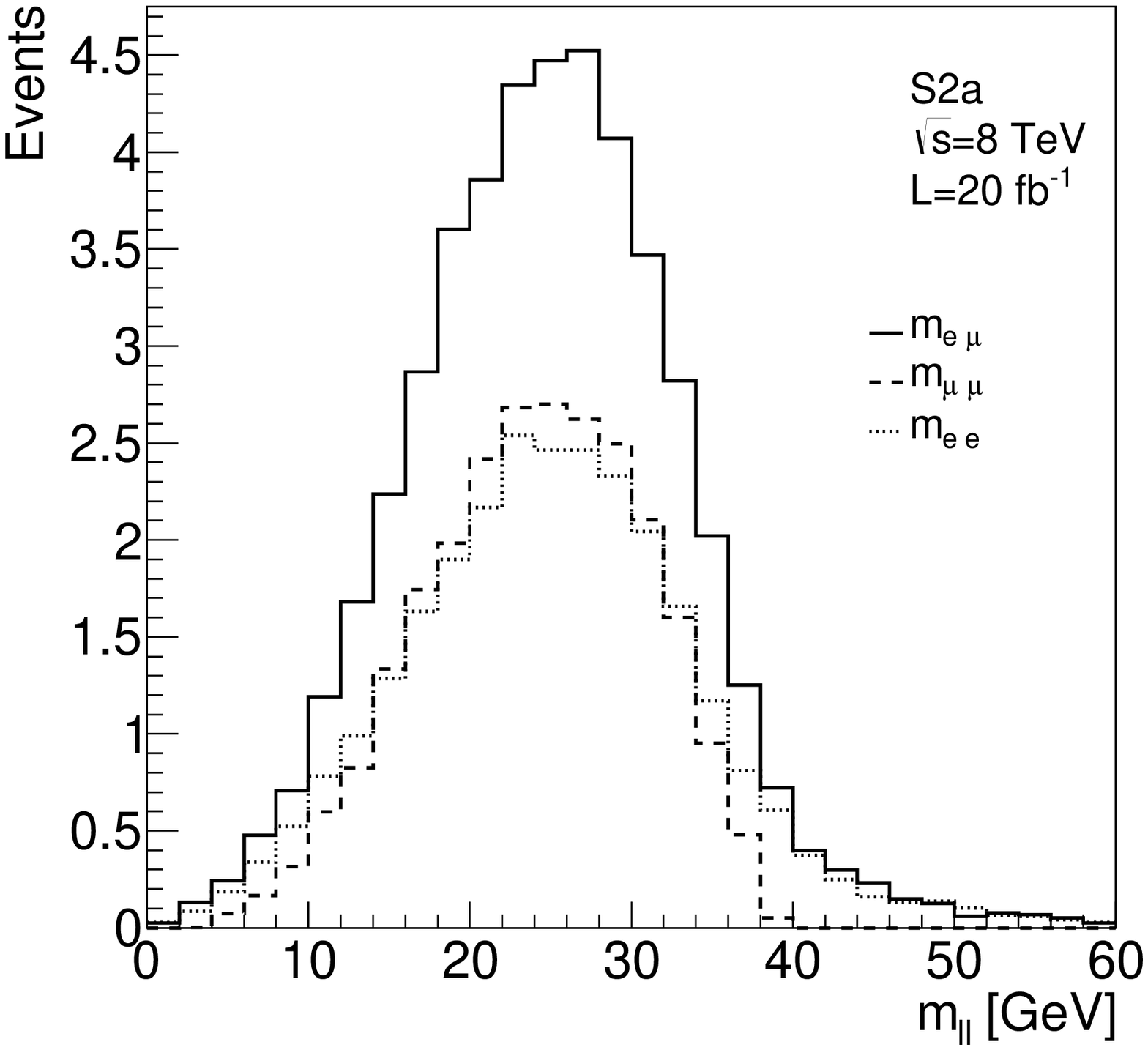}\hspace*{-1ex}
        \includegraphics[scale=0.28]{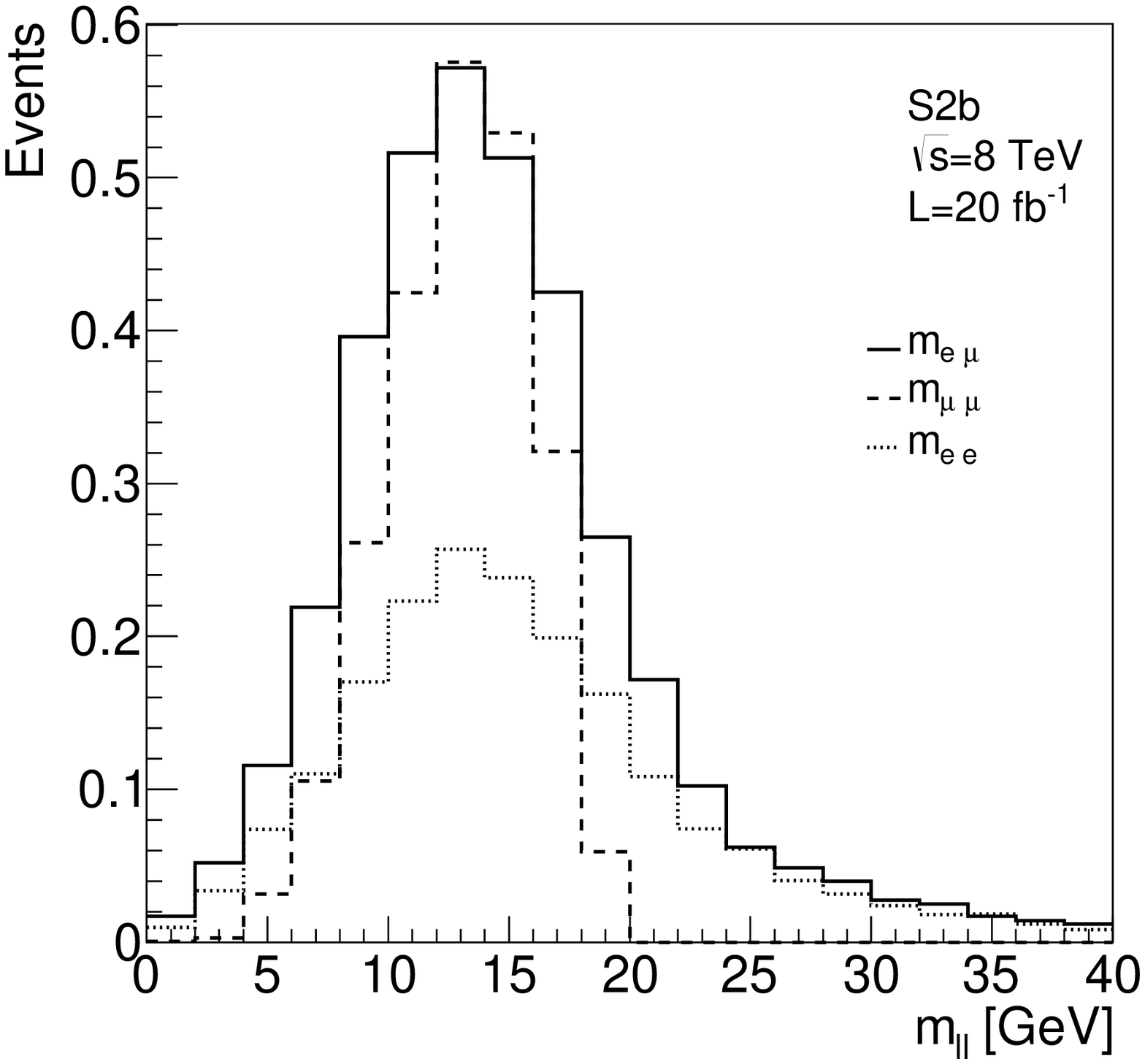}\hspace*{-1ex}
        \includegraphics[scale=0.28]{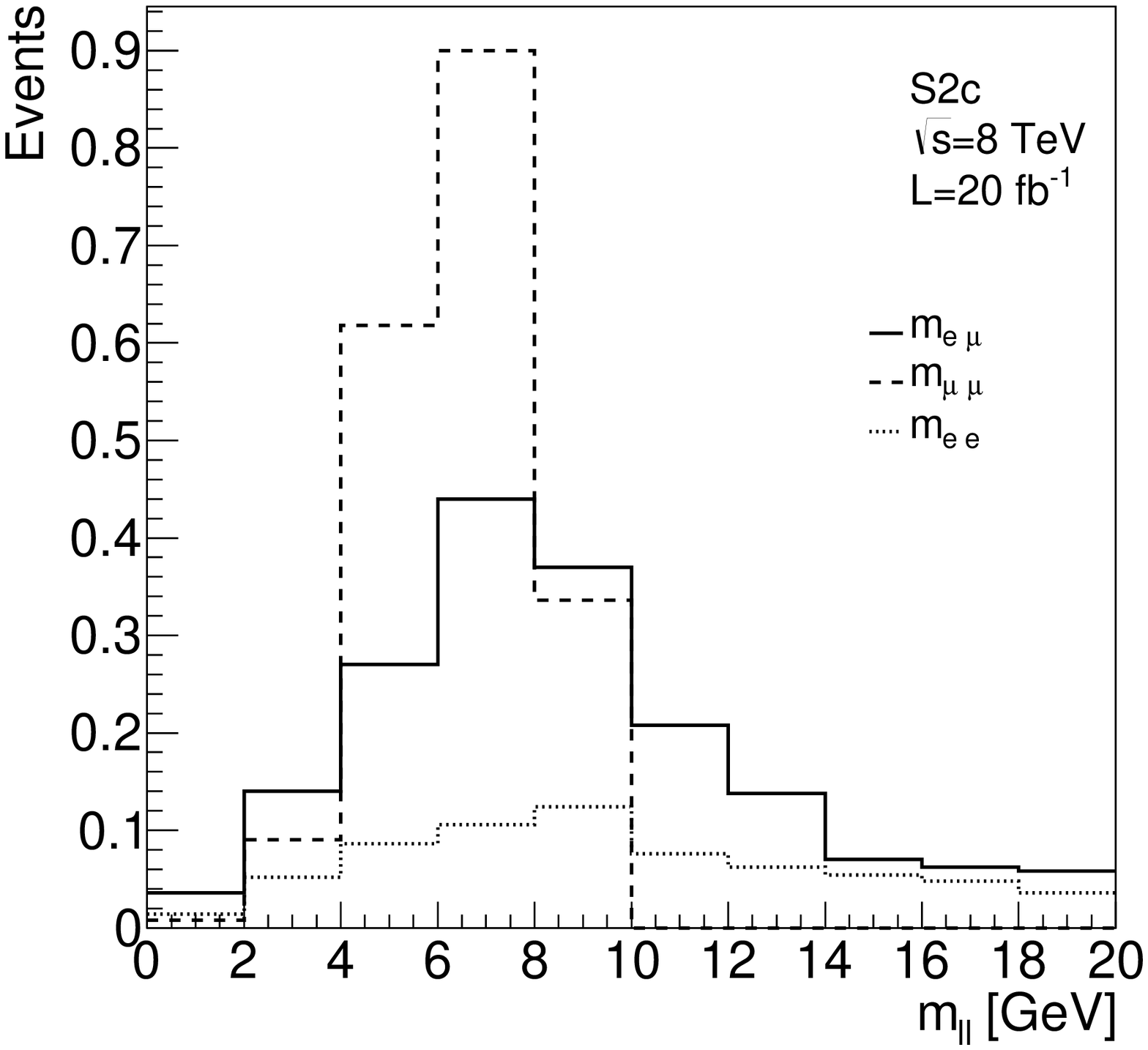}\hspace*{-1ex}
\caption{Two-lepton invariant mass distribution, $m_{\ell\ell}$, for the benchmark points S1a, S1b and S1c (upper row) and S2a, S2b and S2c (lower row) corresponding to the LHC with a centre of mass energy of $\sqrt{s}=8$ TeV and an integrated luminosity of $\mathcal{L}=20$ fb$^{-1}$.
The solid line corresponds to the $m_{e\mu}$, the dashed line represents $m_{\mu\mu}$, and the dotted line is $m_{ee}$.}
\label{fig:mll8}
\end{figure}

Let us first analyse the results obtained for the current LHC configuration, with a centre of mass energy of $\sqrt{s}=8$ TeV and $\mathcal{L}=20$ fb$^{-1}$. We represent in Figure \ref{fig:mll8}  the resulting two-lepton invariant mass distribution for benchmark points S1a, S1b and S1c (upper row) and S2a, S2b and S2c (lower row). The different lines represent the contributions coming from the different channels, $e\mu$ (solid), $\mu\mu$ (dashed), and $ee$ (dotted).
As it was explained above we expect that these distributions present an end-point defined by the kinematics of the system.

In the particular case of S1a, where the $W$ boson is produced on-shell, we can see an edge around $m_{\ell\ell}^{edge}=\sqrt{\rhnmass^2 - m_{W}^2}\simeq 60$ GeV. For this case, the $Z$ peak is present for the same flavour channels. Although this peak is smaller (due to the smaller branching ratio in RH neutrino decays), its observation would allow us to distinguish this distribution from that of a $60$~GeV RH neutrino\footnote{A $60$~GeV RH neutrino would decay through off-shell bosons and present an end-point at its mass.}.
However, for this LHC configuration this peak is actually not observable.

In the rest of the benchmark points the $W$ boson is produced off-shell and the end-point of the invariant mass distribution is at the RH neutrino mass. Nevertheless, the reconstruction of the mass of the RH neutrino could be difficult because the smearing in the lepton energies spoils the tail of the distributions.

Scenarios in which the RH neutrino mass is small, such as S1c, S2b, and S2c are generally difficult to observe since most of the leptons produced fail to pass the cuts on their transverse energy. Also, the smearing on the tail of the dilepton-mass distribution is more severe, due to the small values of the lepton $p_T$. Notice however that case S1c benefits from a sizable production rate and the signal would be very clear.

\begin{figure}[!t]
\hspace*{-4ex}
        \includegraphics[scale=0.28]{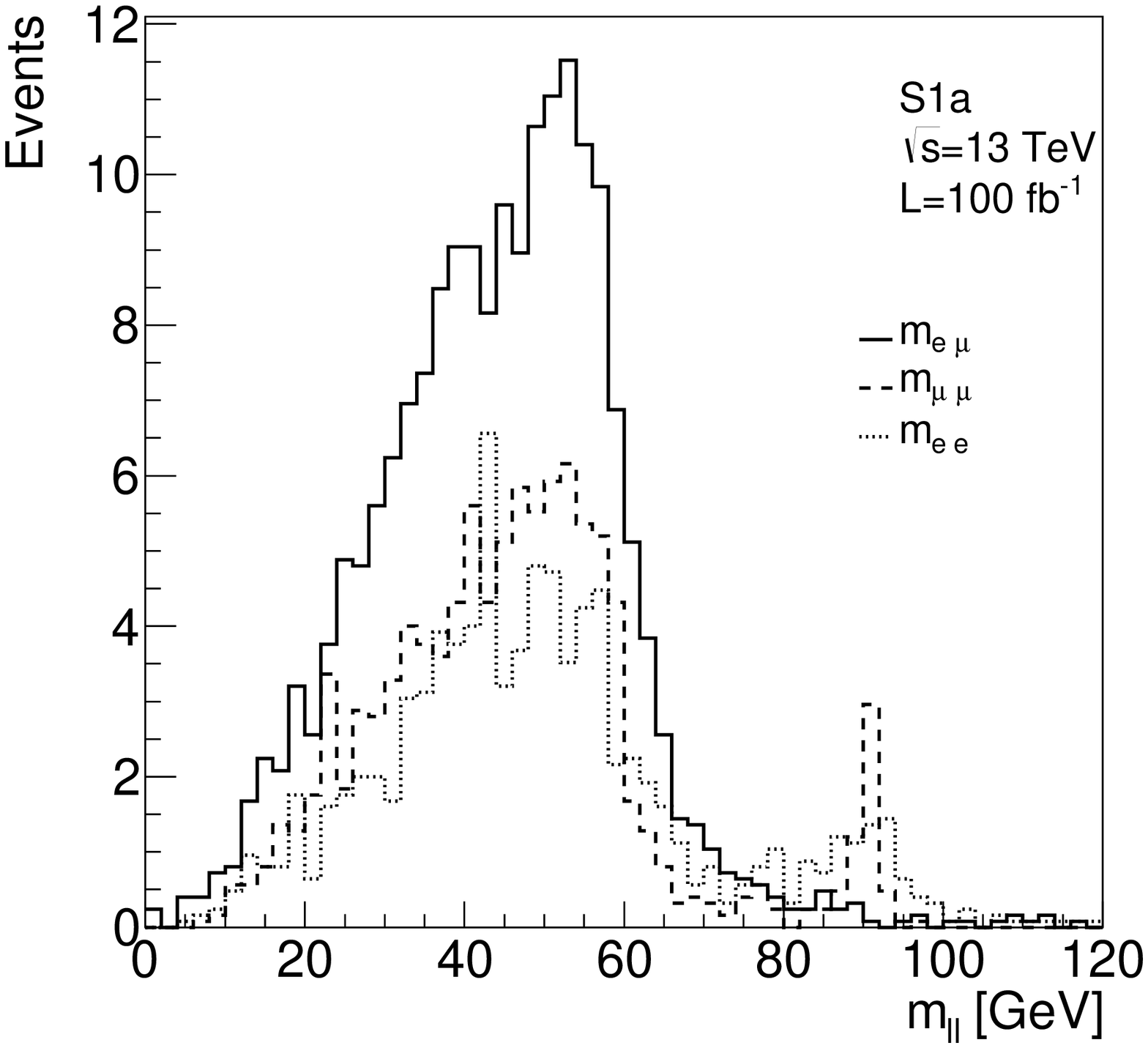}\hspace*{-1ex}
        \includegraphics[scale=0.28]{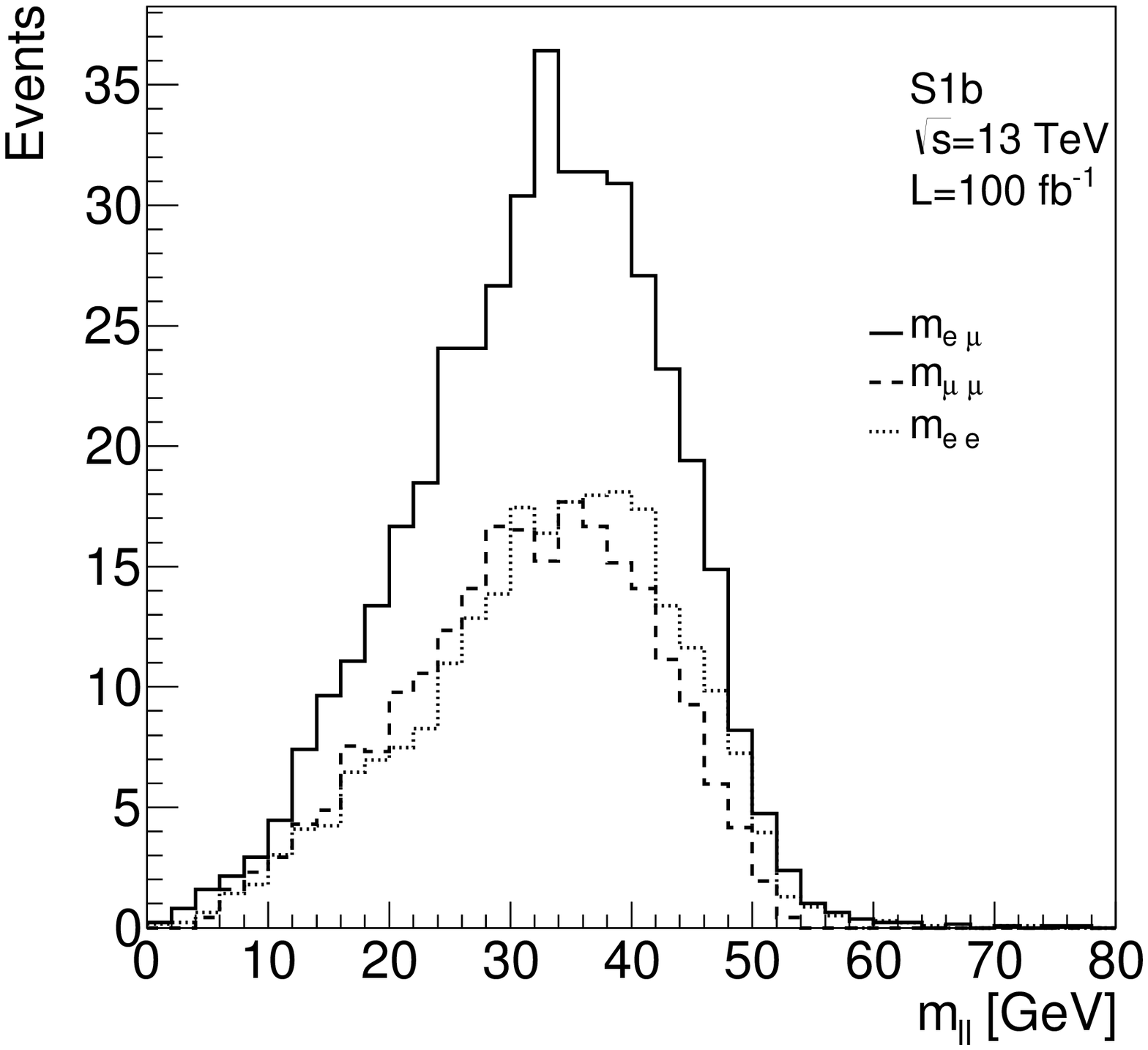}\hspace*{-1ex}
        \includegraphics[scale=0.28]{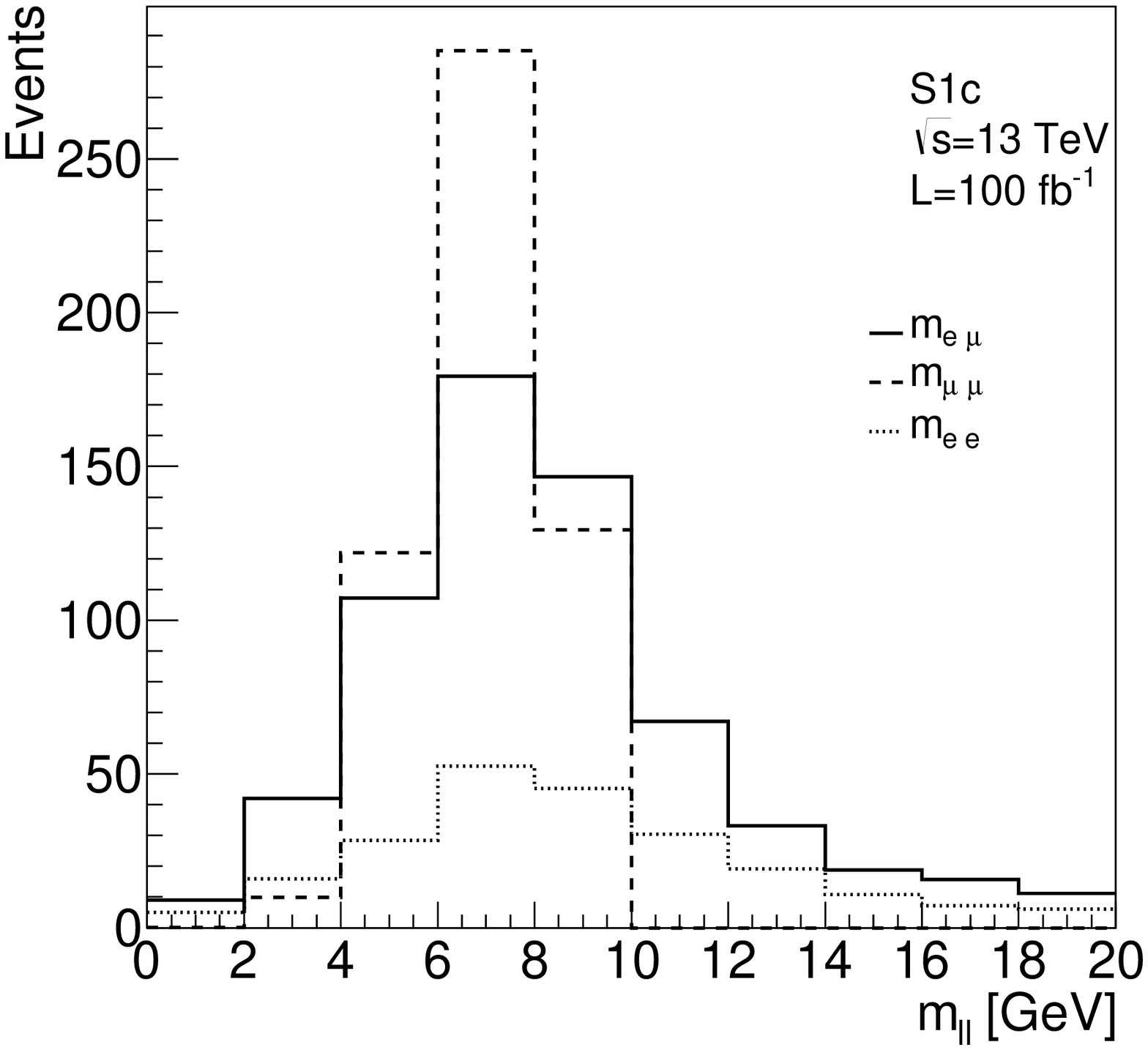}\hspace*{-1ex}\\
\hspace*{-4ex}
        \includegraphics[scale=0.28]{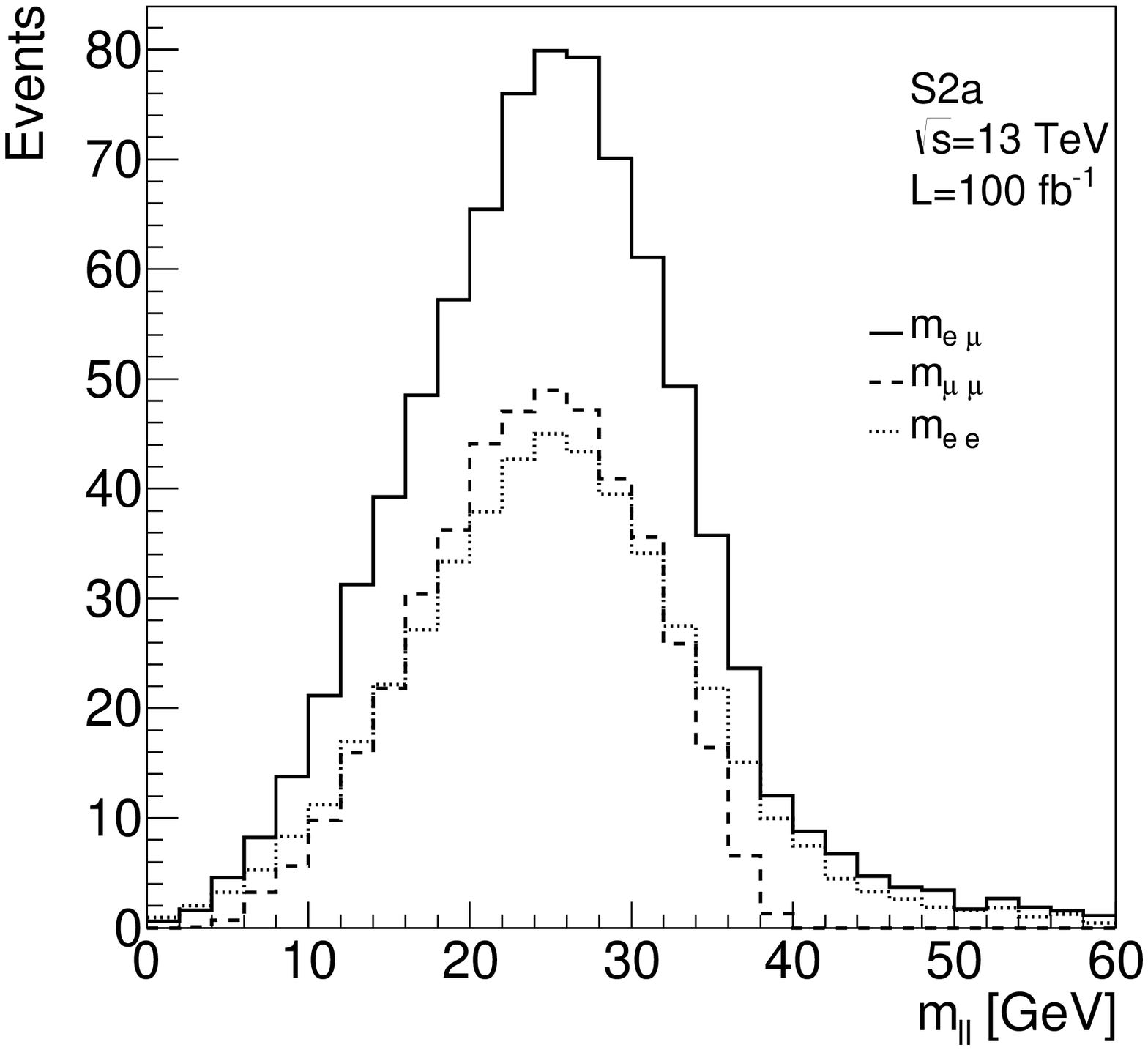}\hspace*{-1ex}
        \includegraphics[scale=0.28]{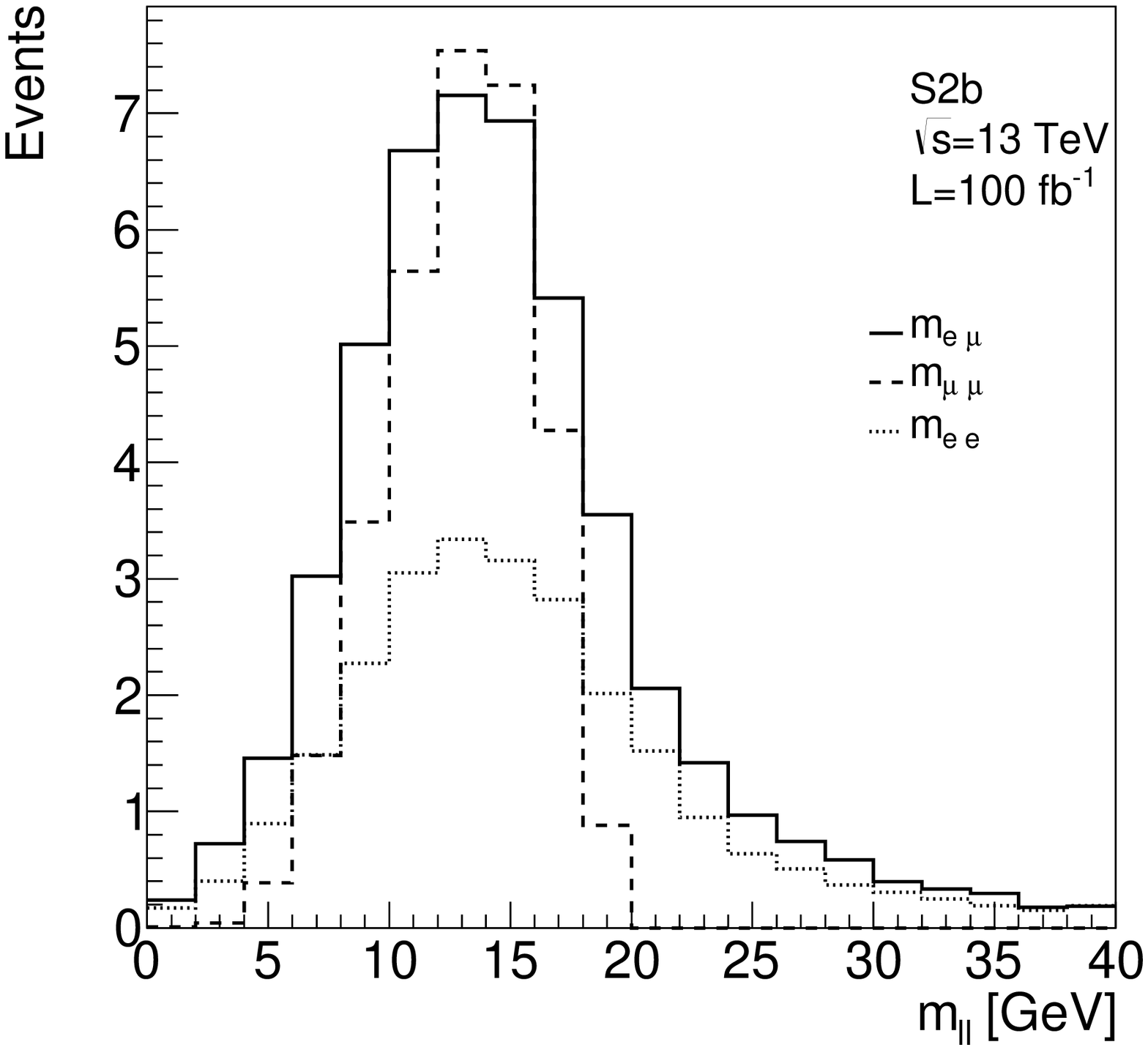}\hspace*{-1ex}
        \includegraphics[scale=0.28]{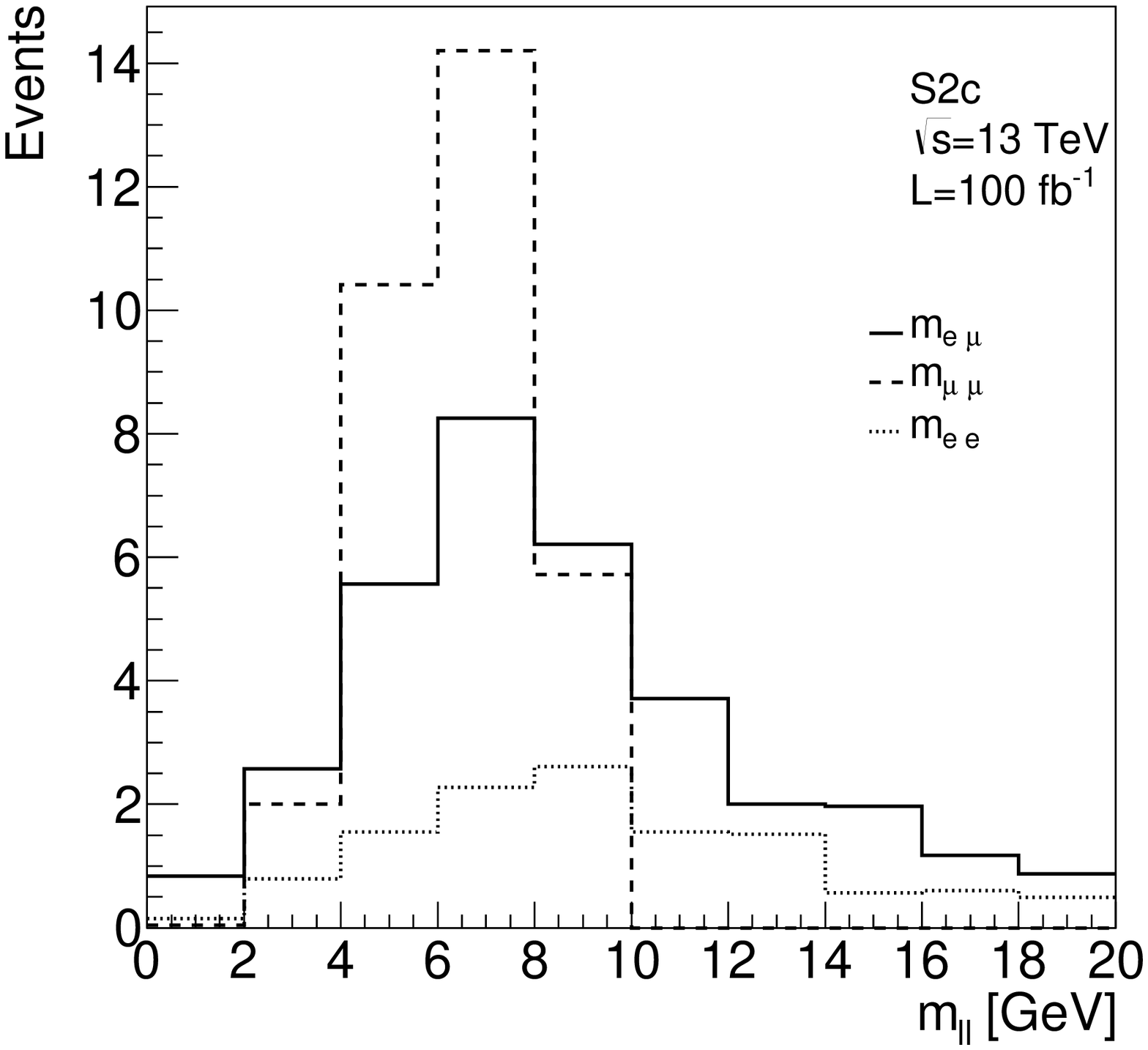}\hspace*{-1ex}
\caption{The same as in Fig.\,\ref{fig:mll8}, but for the LHC with a centre of mass energy of $\sqrt{s}=13$ TeV and an integrated luminosity of $\mathcal{L}=100$ fb$^{-1}$.}
\label{fig:mll13}
\end{figure}

The expected results for the LHC with a centre of mass energy of $\sqrt{s}=13$ TeV and an integrated luminosity of $\mathcal{L}=100$ fb$^{-1}$ are shown in Fig.~\ref{fig:mll13}. The same qualitative results are obtained, but now the number of events is larger and some benchmark points can be probed more easily. 
For example,  the $Z$ peak in benchmark point S1a features 5 events. As this peak is observed in the dimuon channel, we do not expect a depletion in the number of events due to hadronization. Also, since the detector effects are already taken into account through Eq.(\ref{eq:smearing}), we expect that such $Z$ peak would be observable for that scenario in the next configuration of the LHC. 
This is also the case of some examples with low masses, such as S2b and S2c, although the small statistics would make it difficult to determine the end-point of the distributions to extract the RH neutrino mass.

\begin{figure}[!t]
\hspace*{12ex}
        \includegraphics[scale=0.28]{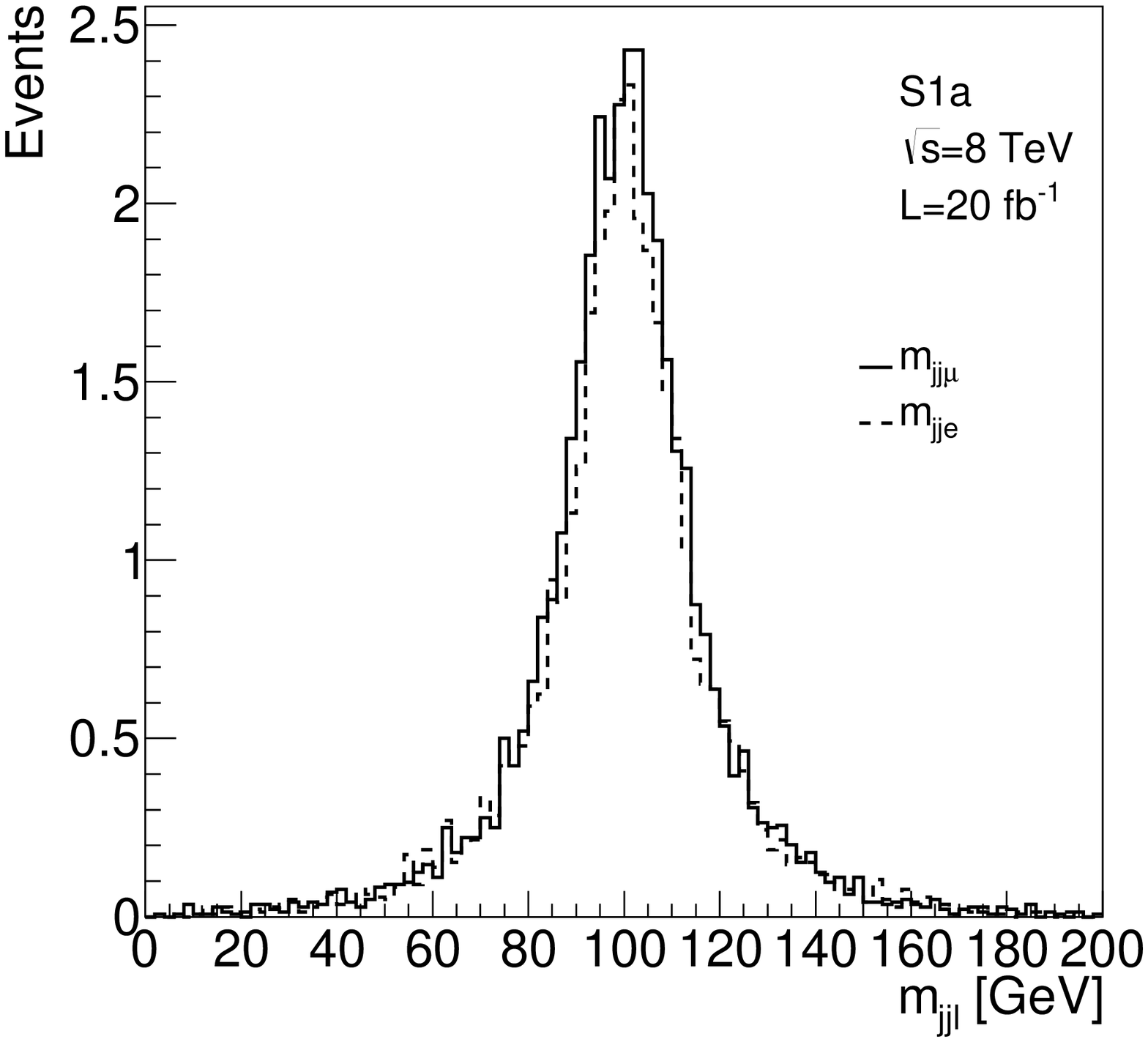}\hspace*{-1ex}
        \includegraphics[scale=0.28]{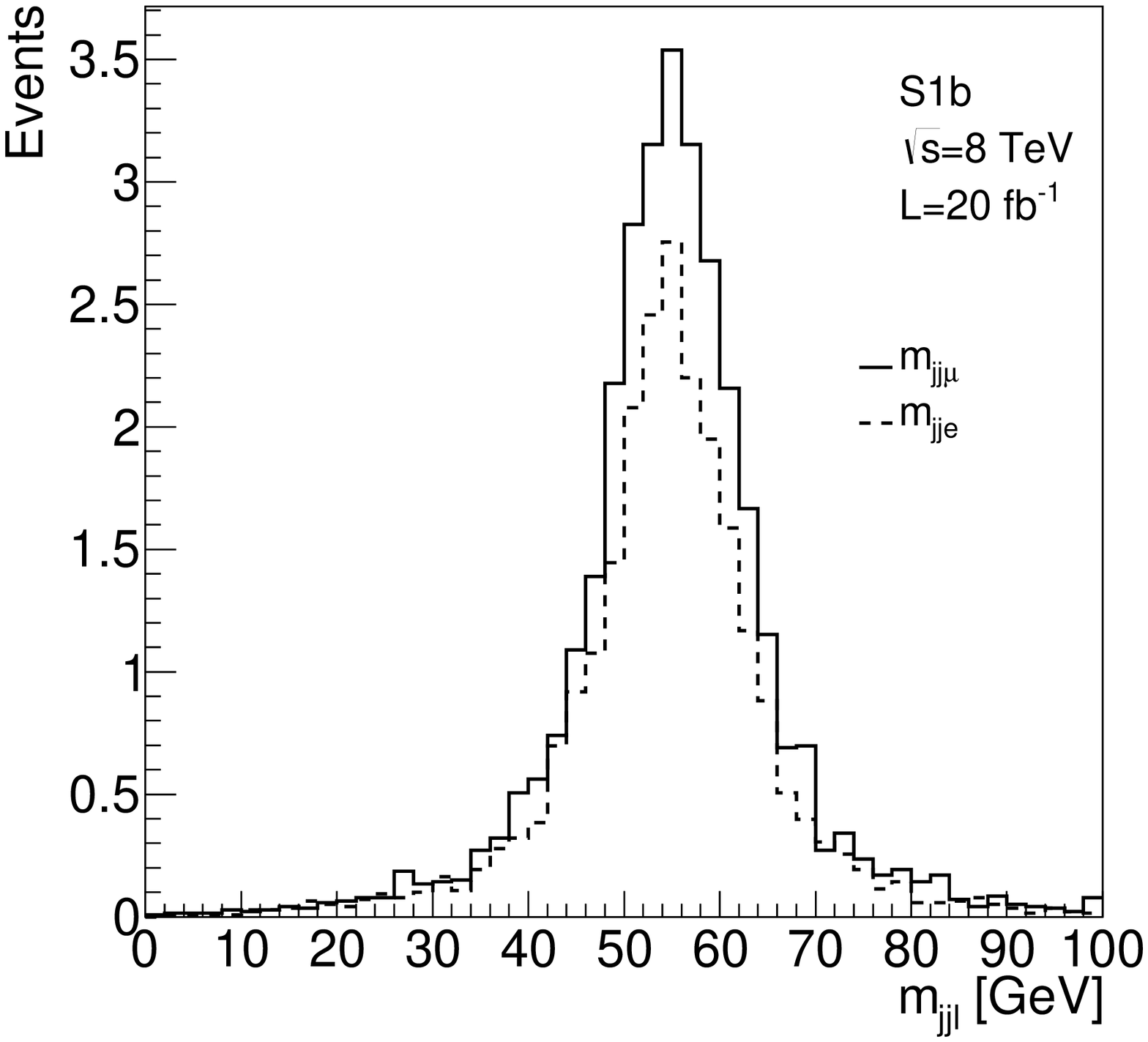}\hspace*{-1ex}
\\
\hspace*{12ex}
        \includegraphics[scale=0.28]{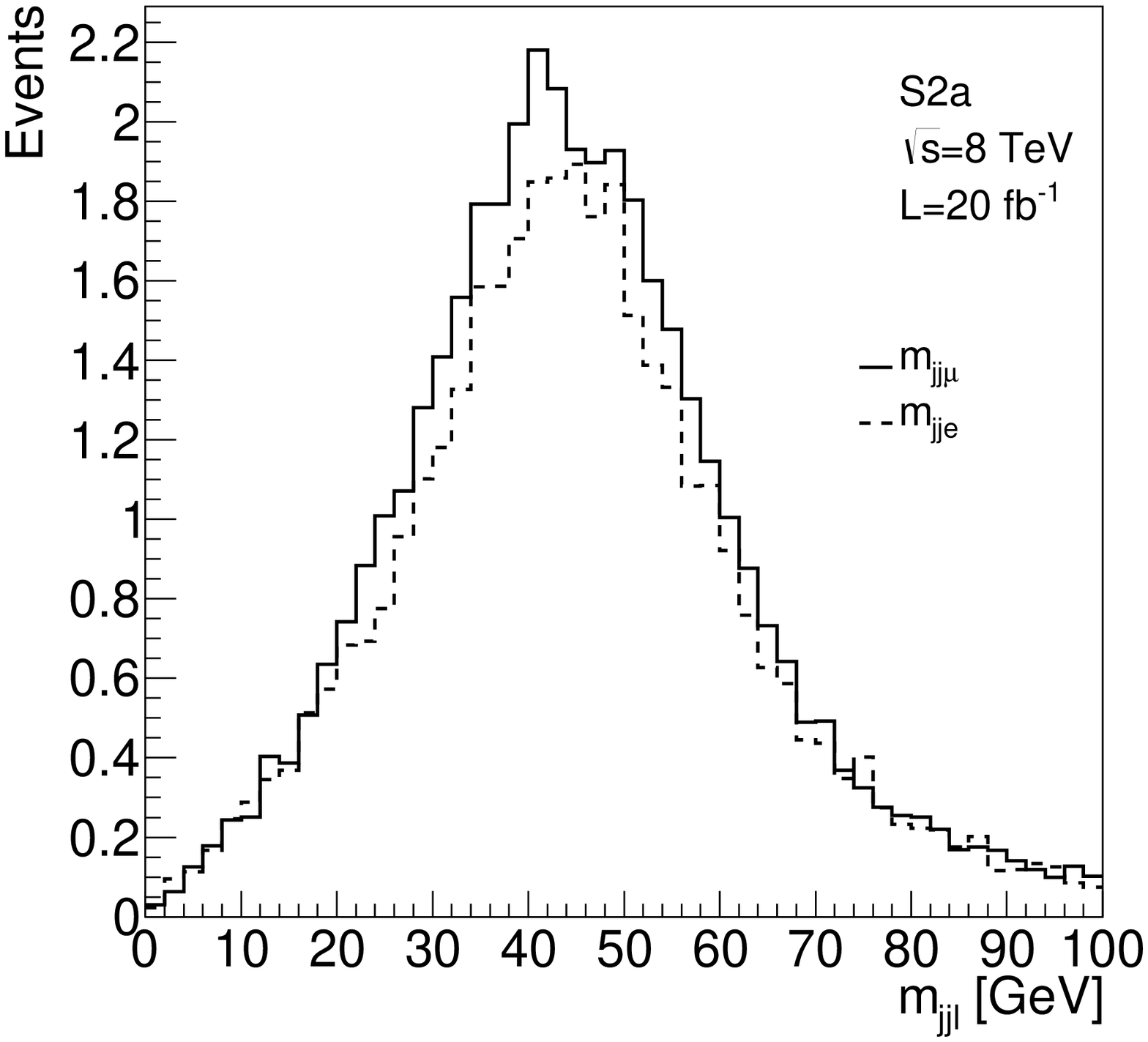}\hspace*{-1ex}
        \includegraphics[scale=0.28]{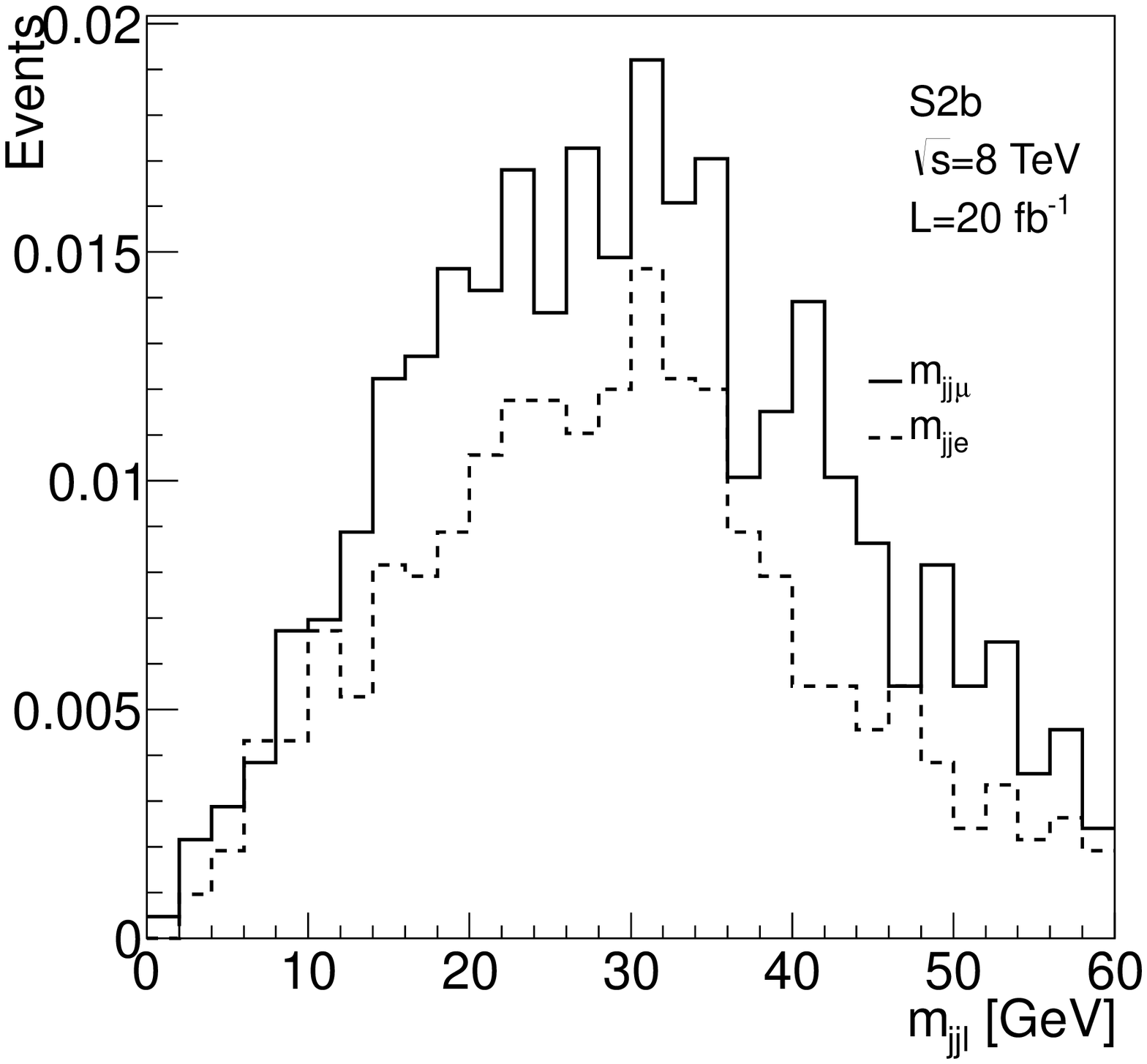}\hspace*{-1ex}
\caption{Invariant mass distribution for two jets and one lepton, $m_{jj\ell}$, for the benchmark points S1a, S1b (upper row), and S2a S2b (lower row) corresponding to the LHC with a centre of mass energy of $\sqrt{s}=8$ TeV and an integrated luminosity of $\mathcal{L}=20$ fb$^{-1}$.
The solid line corresponds to the $m_{jj\mu}$ and the dashed line represents $m_{jje}$.}
\label{fig:mjjl8}
\end{figure}

Let us now turn our attention to the $\rhn\to jj\ell$ signal.  
In Figure \ref{fig:mjjl8} we have represented the two-jets one-lepton invariant mass distribution $m_{jj\ell}$ for the different benchmark points for a LHC configuration of $\sqrt{s}=8$ TeV and an integrated luminosity of $\mathcal{L}=20$ fb$^{-1}$. 
For this distribution, a  peak with a maximum centered in the RH neutrino mass is expected. 
We can see that this is the case in benchmark points S1a, S1b, and S2a. 
From these distributions, the RH neutrino mass can be determined with a certain precision, and compared with the results obtained from the study of the $m_{\ell\ell}$ distribution. 
For S2a we can see that the invariant mass distribution is centered around the mass of the RH neutrino, $\rhnmass=40$ GeV, however the width of the distribution is larger. Although the jets can pass the cuts, they have a small energy and cannot be reconstructed properly due to the smearing effects.

If the RH neutrino mass is small (as in benchmark points S1c, S2b, and S2c), the jets are less energetic and are more affected by the cut in $p_T$.
For these three benchmark points, the jets and leptons cannot fulfill the cut requirements and no events would be observed (see Table~\ref{tab:events}).

\begin{figure}[!t]
\hspace*{12ex}
        \includegraphics[scale=0.28]{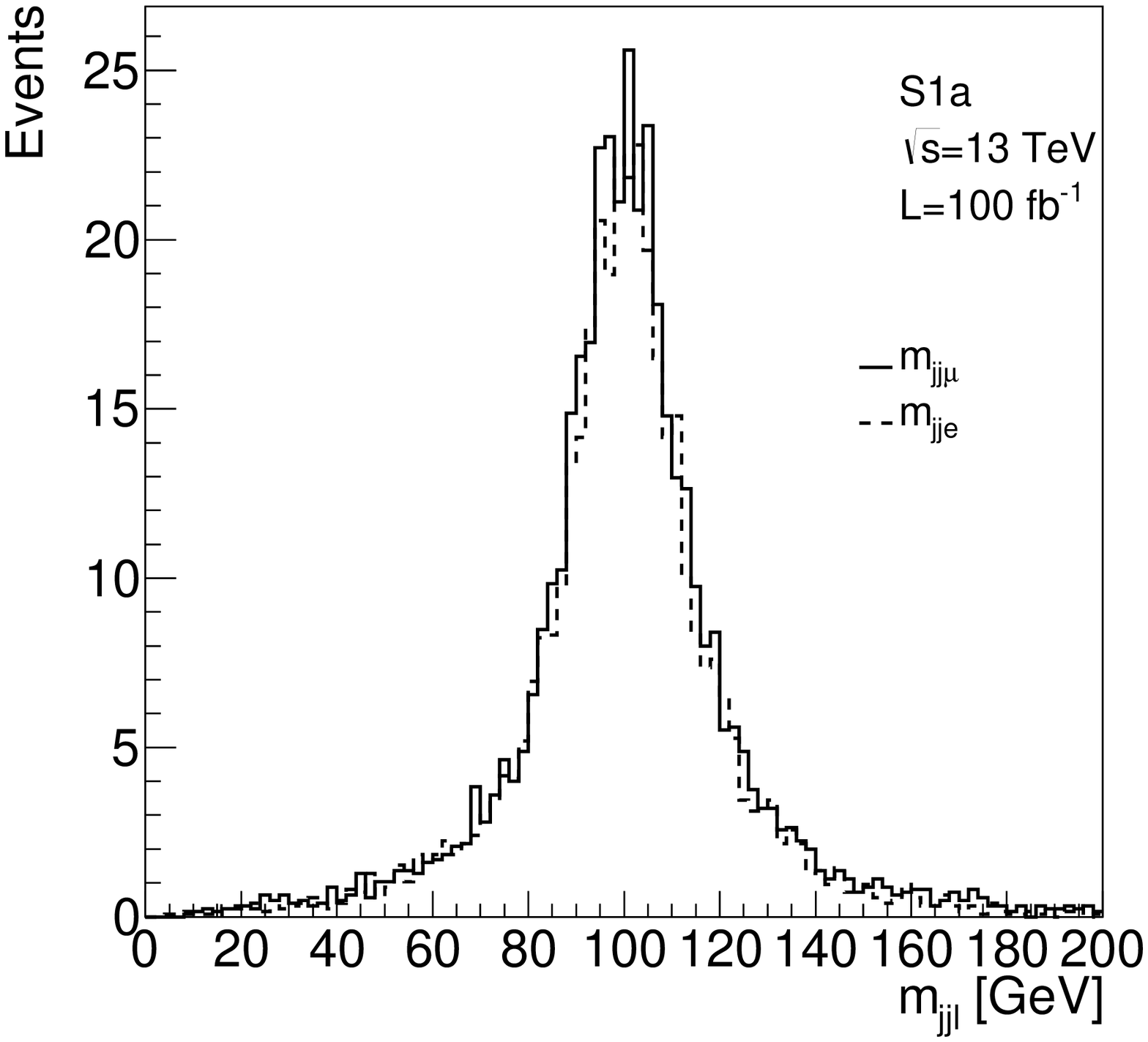}\hspace*{-1ex}
        \includegraphics[scale=0.28]{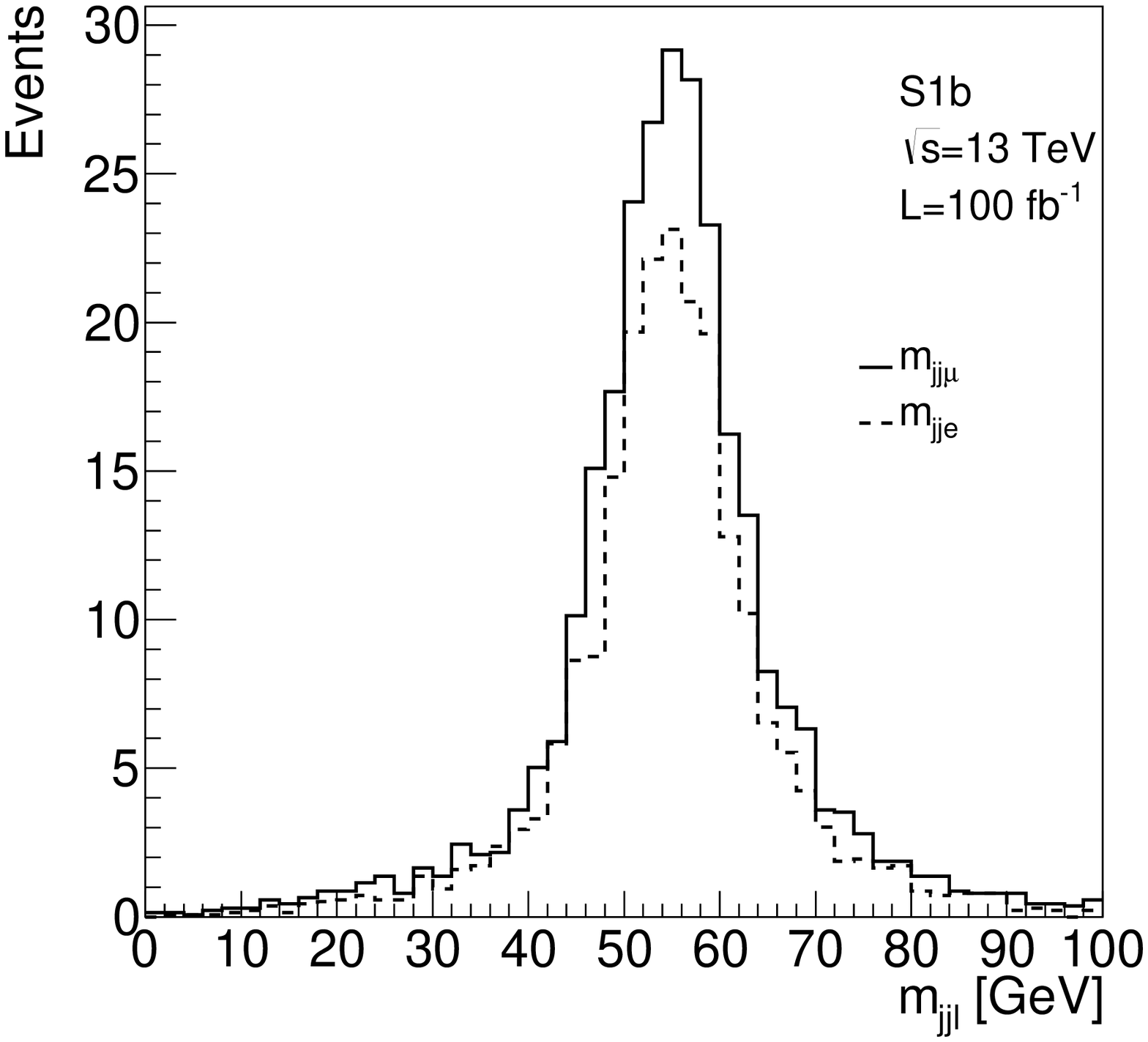}\hspace*{-1ex}
\\
\hspace*{12ex}
        \includegraphics[scale=0.28]{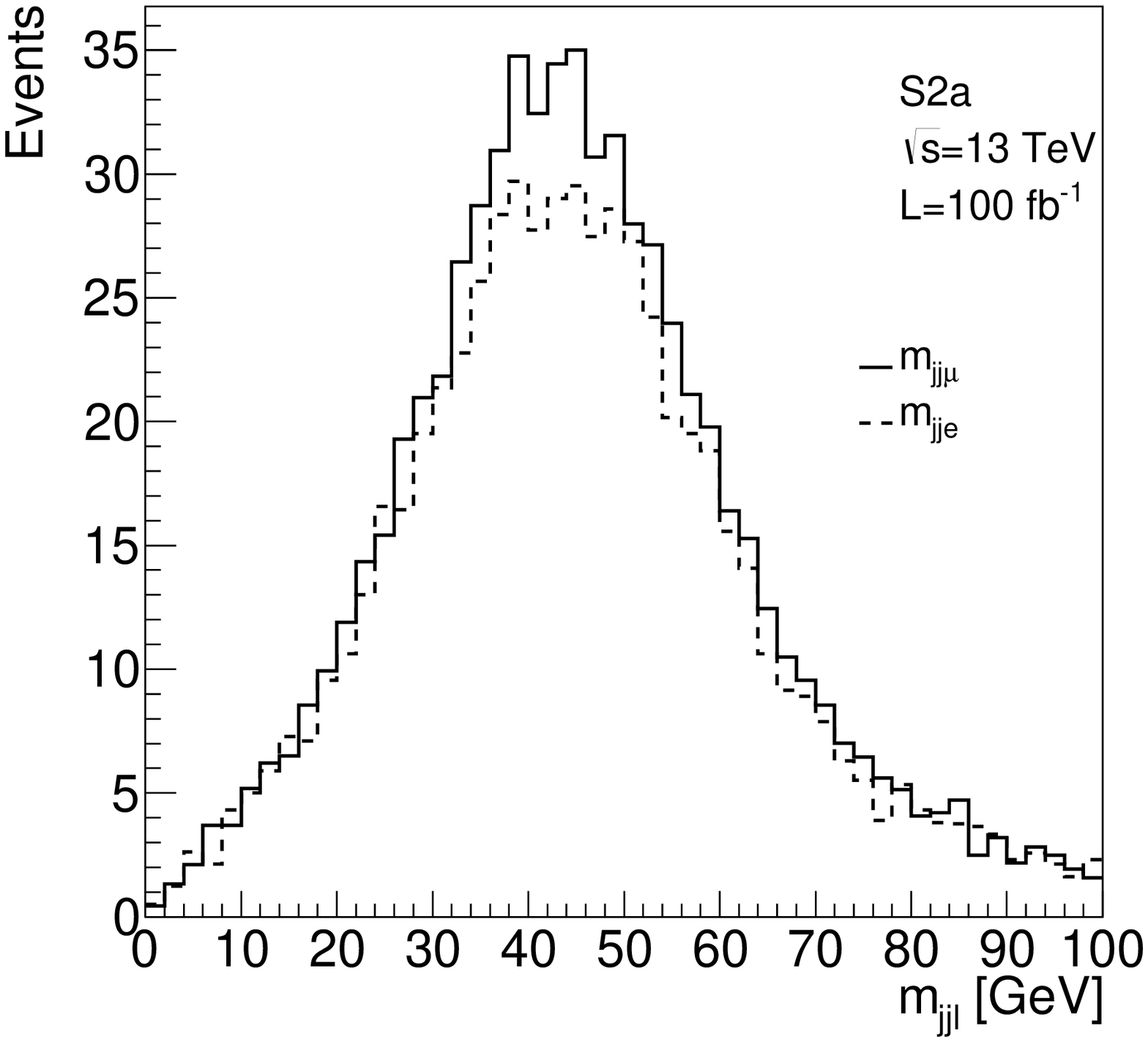}\hspace*{-1ex}
        \includegraphics[scale=0.28]{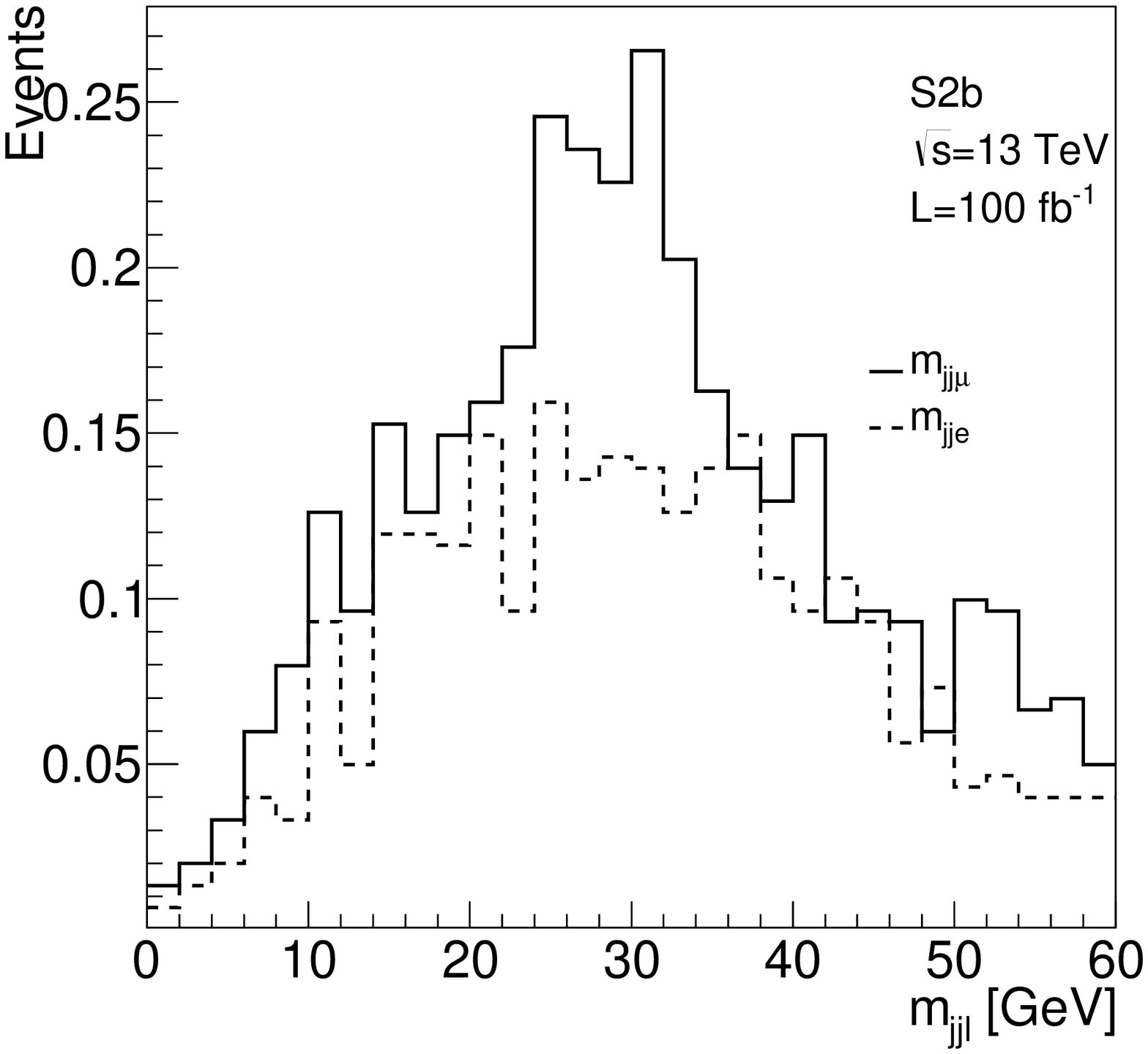}\hspace*{-1ex}
\caption{The same as in Fig.\,\ref{fig:mjjl8}, but for the LHC with a centre of mass energy of $\sqrt{s}=13$~TeV and an integrated luminosity of $\mathcal{L}=100$ fb$^{-1}$.}
\label{fig:mjjl13}
\end{figure}

If we now consider the future LHC configuration, with a centre of mass energy of $\sqrt{s}=13$ TeV and an integrated luminosity of $\mathcal{L}=100$ fb$^{-1}$, the number of events increases and the reconstruction of the RH neutrino mass is clearer. We show the corresponding distributions of $m_{jj\ell}$ in Fig.~\ref{fig:mjjl13}. We can observe that the reconstruction for the benchmarks in scenario S2 is less precise, as explained above, due to the smaller energy of the resulting jets. Benchmark point S2b is now observable (with approximately 12 events), however both S1c and S2c remain unobservable and are therefore not shown.

Notice that the results from Refs.\cite{ATLAS:2012av, Aad:2012kw, Aad:2012zx, Chatrchyan:2012jna} are the present constraints from the LHC on displaced vertices. Some of these searches share the same signatures with this model. As these searches are done in the $\sqrt{s}=7$ TeV with luminosities less than $\mathcal{L}=5$ fb$^{-1}$ and they impose strong cuts in the $p_T$ of the objects that arise from the displaced vertex we found that our benchmark points agree with the lack of signals that these searches found.

Also, due to the fact that some of the RH neutrinos could decay promptly, the decay objects could contribute to multilepton signals in standard ATLAS and CMS searches for supersymmetry \cite{ATLAS:multilepton, CMS:multilepton}. We have simulated the expected number of multilepton events coming from RH neutrino decays with an impact parameter smaller than $|d_0|< 0.2$ mm, and observed that this number is smaller than one in all the benchmark points. This means that the present searches on multilepton signals do not constrain our scenarios.

It should finally be mentioned that displaced vertices can also appear in R-parity violating (RPV) supersymmetric models~\cite{Barbier:2004ez}.
For example, this is the case for a realization of these scenarios with trilinear RPV through
a $\lambda'' UDD$ term in the superpotential~\cite{Carpenter:2007zz}
can induce displaced vertices~\cite{DreinerRoss}.
However, the final states in these RPV models are different to the ones observed in our scenario, as they originate from different couplings.
In particular, the $LLE$ operator leads only to $2\ell+\met$, 
the $LQD$ operator leads only to  $\ell jj$ and $jj+\met$, and the $UDD$ operator leads only to $jjj$.

Similarly, bilinear RPV models with
 $\Delta W = \mu_i L_i H_u$, 
can also account for non-vanishing neutrino masses through the neutralino-neutrino mixing.
The final state produced at the displaced vertex in these scenarios from the decay of unstable neutralinos would be
 $2\ell+ \met$ and $jj+\met$ by $\nu_L)$, where in both cases the missing energy is due to the production of a $\nu_L$~\cite{Graham:2012th}, however we would not observe any $\ell jj$ events.

Contrary to trilinear and bilinear RPV, in our scenario the  $2\ell+\met$ and $\ell jj$ signatures have the same origin (the decay of the long-lived $N$). As we have shown, from the reconstructed end-point in the two-lepton invariant mass distribution ($m_{\ell\ell}$) and the peak in the two-jets one lepton invariant mass distribution ($m_{\ell j j }$) we would reconstruct the same value of the RH neutrino mass. This is a valuable cross-check that would allow us to discriminate our scenario from the above mentioned RPV models.

\section{Long-lived charged particles}
\label{sec:stau}

A charged and long-lived particle can leave a distinctive track at the LHC that could be identified as corresponding to a particle heavier than a muon.
In our construction, this can be the case, for example, of the lighter stau, which eventually decays into the RH sneutrino.

There are various contributions to the stau decay, depending on its mass:
\begin{itemize}
\item $\tilde{\tau}_1\to W{\snr_1}$

This is the only two-body decay channel which is kinematically allowed when $m_{\tilde{\tau}}\ge m_W + \snmassr$. It is suppressed by the mixing in the sneutrino sector, which is proportional to $y_N$.

\item $\tilde{\tau}_1\to q_i \bar{q}_j\snr_1$, $\nu_L l\snr_1$

These processes are mediated by 
 a virtual  $W$ boson that connects to a $q_i\bar q_j$ pair or $\nu_L l$. As in the former example, the sneutrino arises through the mixing with $\tilde\nu_L$,  which is proportional to $\yn$. 

\item $\tilde{\tau}_1\to \tau \rhn\snr_1$

This process is mediated by a neutralino $\chi_i^0$ and is not Yukawa suppressed.

\end{itemize}

As in the case of the neutralino NLSP, the first two channels include a dependence on the neutrino Yukawa through the mixing of the RH sneutrino with the LH ones, and this implies a small decay width and a long lifetime, which in general would allow the stau to escape the detector.
Notice however that the third channel is not Yukawa suppressed and therefore dominates when it is kinematically allowed (when $m_{\tilde{\tau}_1}> m_{\tau}+ \rhnmass+\snmassr$).

\begin{figure}[!t]
        \includegraphics[scale=0.405]{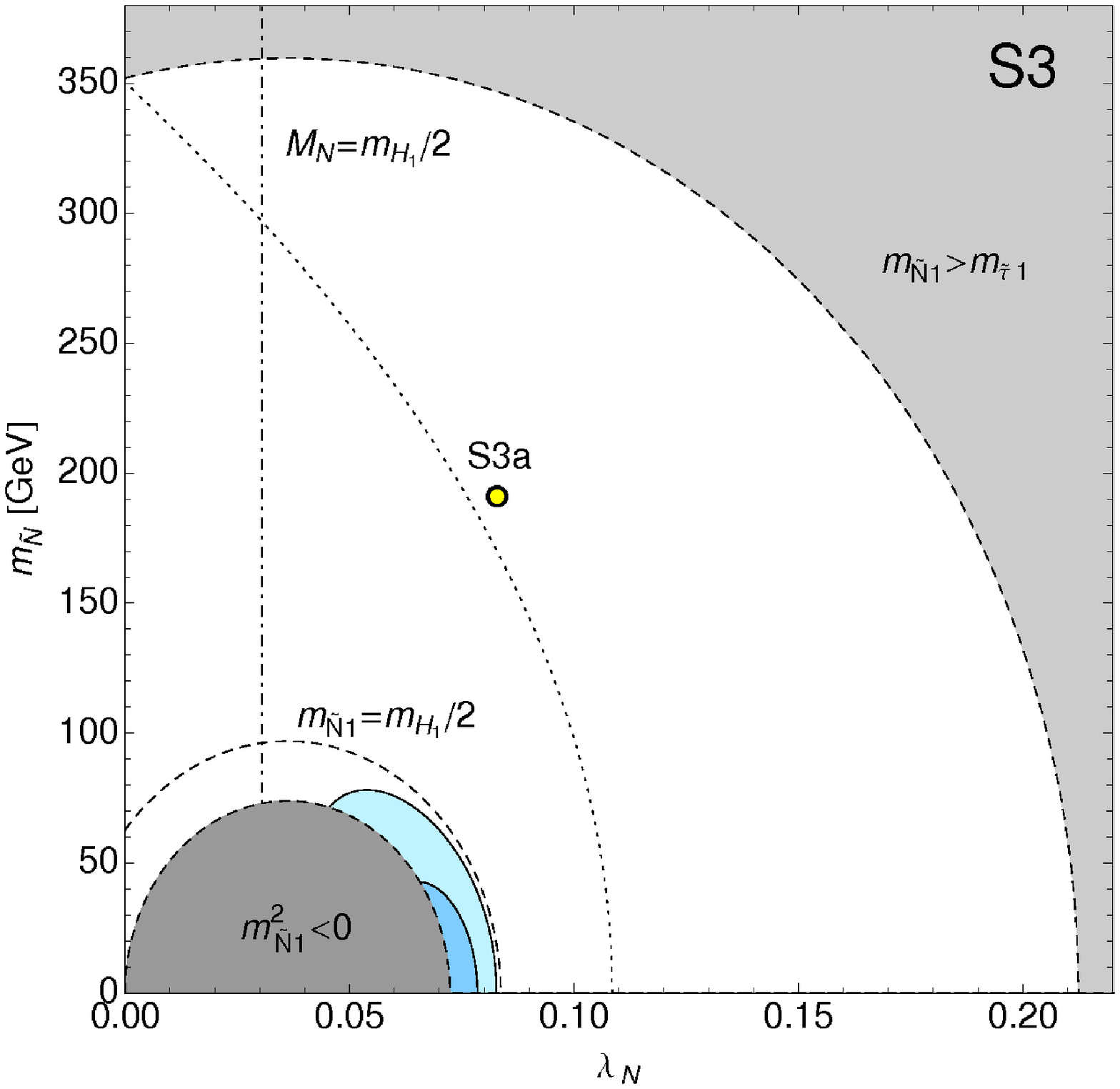}\hspace*{1cm}
        \includegraphics[scale=0.405]{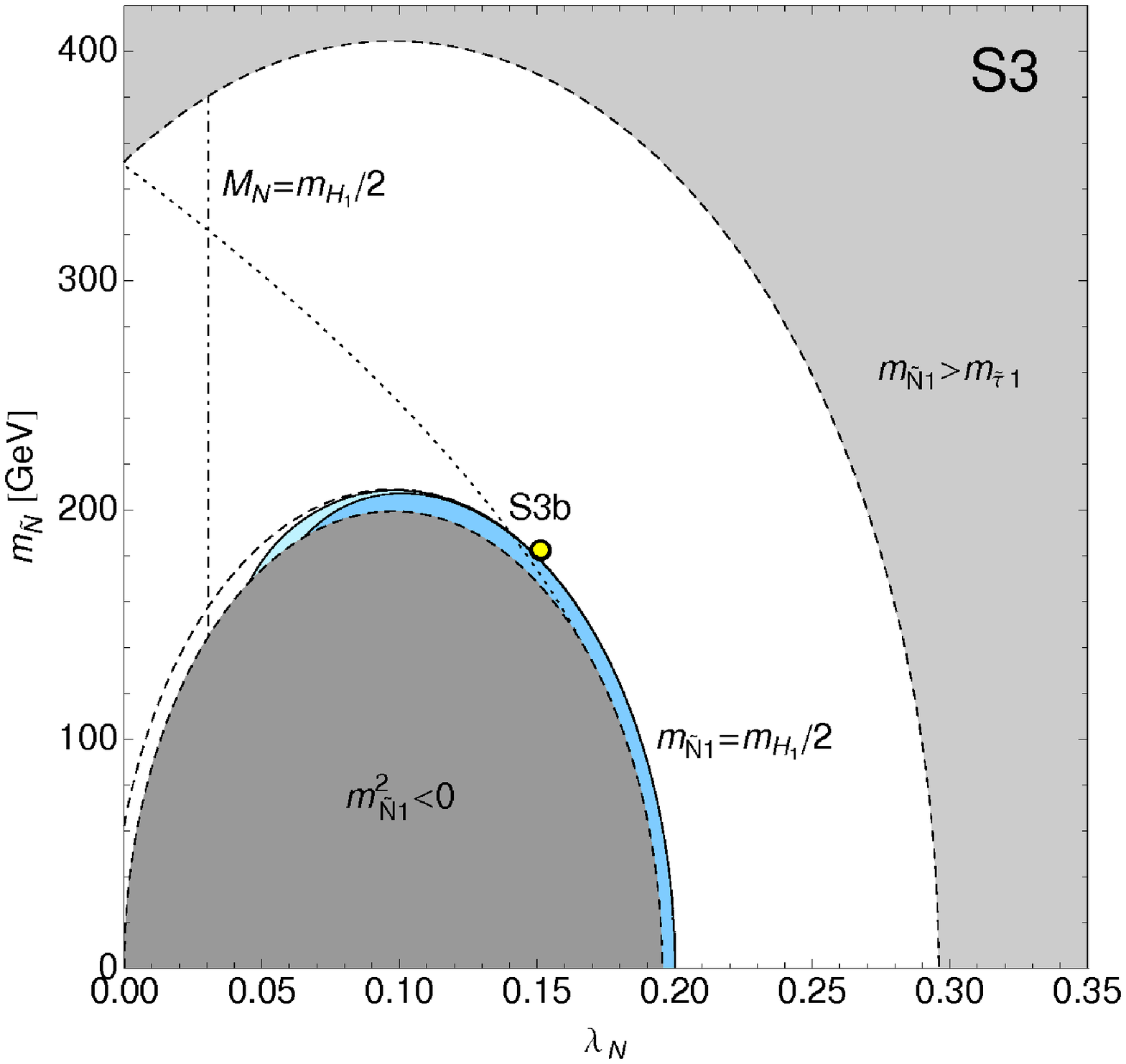}
        \caption{\small
Constraints on the $(\ln,\,\mn)$ plane from the invisible branching ratio of the SM-like Higgs for scenario S3. From left to right, the  trilinear term is $\aln=-500$, and $-750$~GeV. The same colours and lines as in Fig.\,\ref{fig:invisible} are used. Points to the left and below the dotted line satisfy $m_{\tilde{\tau}_1}> m_{\tau}+ \rhnmass+\snmassr$. Yellow dots correspond to the selected benchmark points.}
\label{fig:invisibles3}
\end{figure}

For concreteness, we will study scenario S3 in Table \ref{tab:S-12}, which features a stau NLSP. In Fig.\,\ref{fig:invisibles3} we represent the corresponding $(\ln,\,\mn)$ plane for two choices of the trilinear parameter $\aln=-500$ and $-750$~GeV and indicate the areas that are excluded by the constraint on the invisible Higgs decay. As in scenarios S1 and S2, wide regions of the parameter space are available.
Points to the left and below the dotted line satisfy $m_{\tilde{\tau}_1}> m_{\tau}+ \rhnmass+\snmassr$ and correspond to areas in which the stau can decay promptly.

\begin{figure}[!t]
\begin{center}
        \includegraphics[scale=0.39]{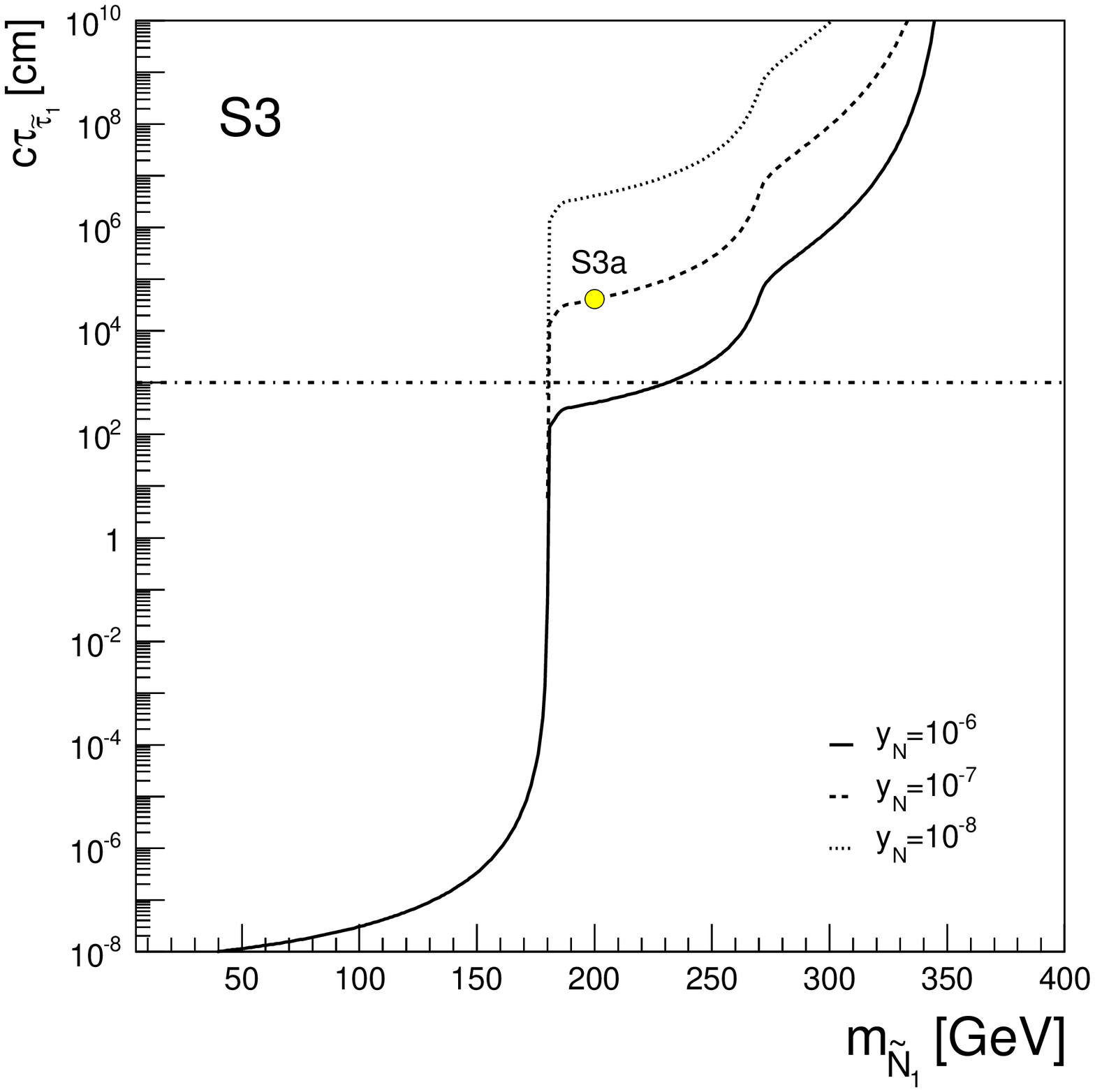}\hspace*{1.ex}
        \includegraphics[scale=0.39]{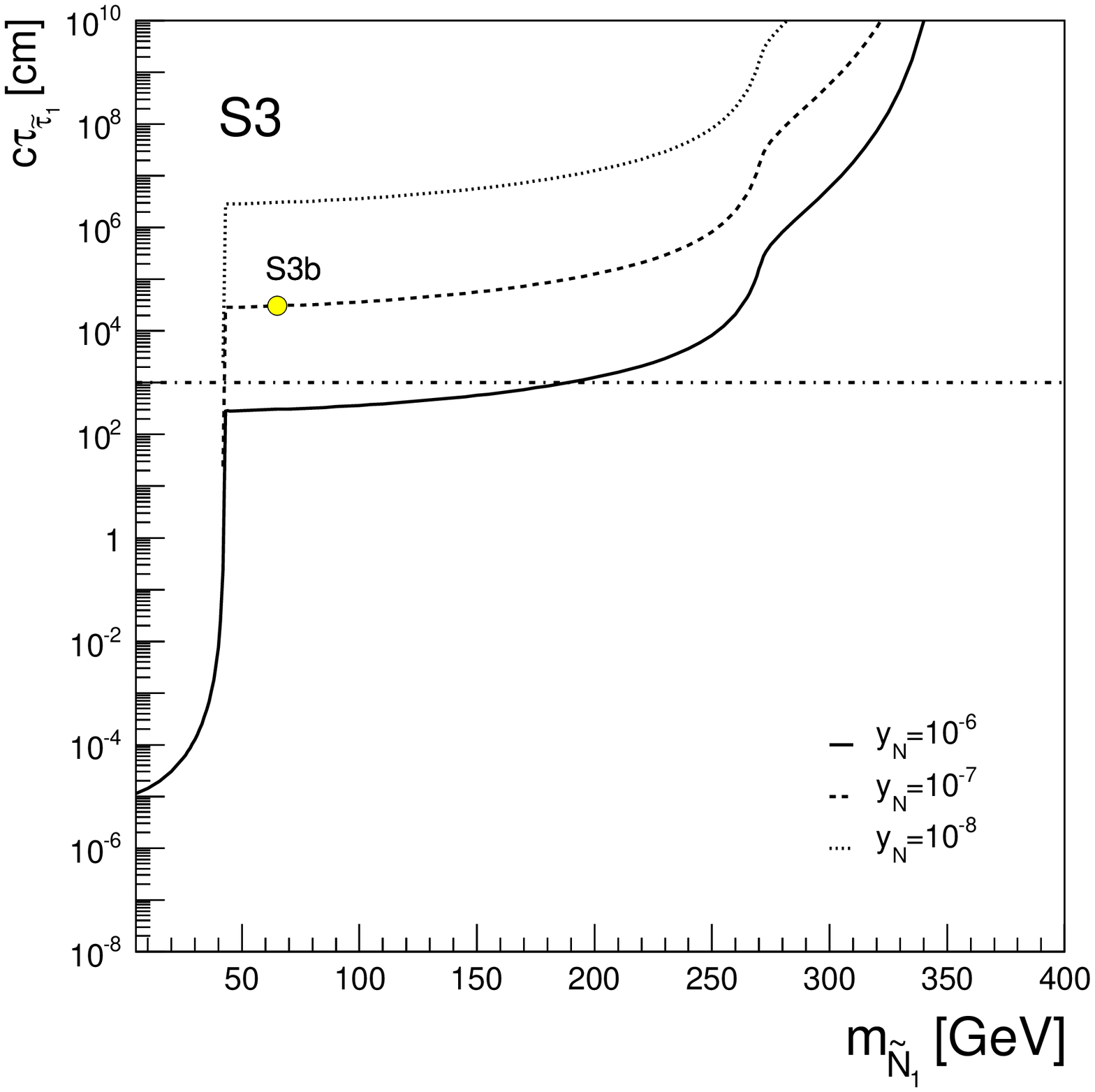}
\caption{Decay length of the lighter stau NLSP as a function of the RH sneutrino mass for scenario S3 with a fixed RH neutrino mass of $170$~GeV (left) and $310$~GeV (right). The different lines represents different values of the neutrino Yukawa coupling. The dot-dashed line at $c\tau=10^3$~cm represents the approximate radius of the ATLAS detector. Yellow circles denote benchmark points S3a and S3b.}
\label{fig:ctaustau}
\end{center}
\end{figure}

We have computed the different contributions to the stau lifetime for two examples, based on scenario S3, where the RH neutrino mass has been fixed to $170$ and $310$~GeV. The results are represented in Fig.\,\ref{fig:ctaustau} as a function of the RH sneutrino mass. We observe that the stau decays outside the detector for the whole range of relevant values of the neutrino Yukawa, $\yn\approx10^{-6}-10^{-8}$, and RH sneutrino masses, except for the region with a light RH sneutrino for which the $\tilde{\tau}_1\to \tau \rhn\snr_1$ decay is kinematically open\footnote{For $\yn=10^{-6}$ the stau can decay inside the detector for small RH sneutrino masses. However, the decay takes place in the calorimeter and not in the inner detector. We consider this possibility difficult to identify and do not consider it in the analysis. }. 
We have chosen two benchmark points, S3a and S3b, with a stau mass $m_{\tilde{\tau}_1}=352$~GeV and parameters defined in Table \ref{tab:S-12}. The stau lifetime for both is represented by yellow circles in Figs.\,\ref{fig:invisibles3} and \ref{fig:ctaustau}.

\begin{figure}[!t]
\begin{center}
        \includegraphics[scale=0.85]{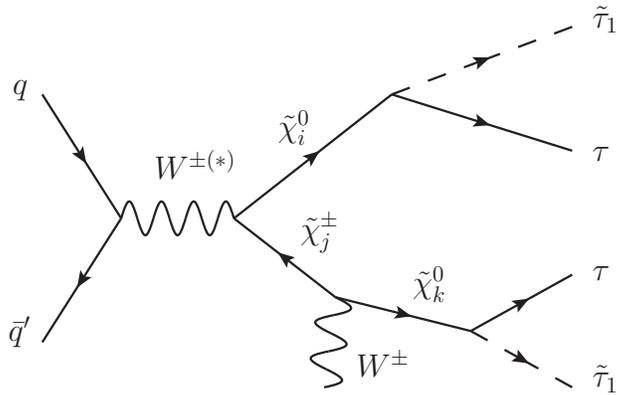}
\caption{Long-lived stau production considered in this model. A neutralino/chargino pair is produced after the original collision and undergoes a short decay chain that ends in the production of long-lived staus and tau leptons. }
\label{fig:prodstau}
\end{center}
\end{figure}

For each example we have simulated the production of long-lived staus in proton-proton collisions. The main production of the stau NLSP comes from the decay chains originated after the creation of neutralino/chargino pairs as illustrated in Fig.\,\ref{fig:prodstau}. We consider the current LHC configuration with a centre of mass energy of 8 TeV and an integrated luminosity of $\mathcal{L}=20\, \text{fb}^{-1}$, and the future one, with a centre of mass energy of 13 TeV and $\mathcal{L}=100\, \text{fb}^{-1}$. The total neutralino/chargino production cross sections for each centre of mass energy ($\sigma^{8,\, 13\,\text{TeV}}_{\tilde{\chi}_j^\pm\tilde{\chi}_i^0}$) is written in Table\,\ref{tab:eventss3}.
In both benchmark points the lighter neutralino decays as $\neut_1\to\tau\tilde{\tau}_1$ with a branching ratio which is approximately 100\% (notice that since we have chosen a heavy RH neutrino, the direct decay $\neut_1\to \rhn\snr_1$ is kinematically forbidden and $\neutl\to\snr\nu_L$ is suppressed by $\yn$).

We impose the following basic cuts, aimed at reducing the background  (mostly due to high $p_T$ muons) \cite{ATLAS:longlivedsleptons}.
\begin{itemize}
\item[-] We require {\em two} staus which escape the detector ($c\tau>10\,{\rm m})$.
\item[-] In order to discriminate heavy long-lived staus from muons, the measured $\beta\equiv v/c$ is required to be less than 0.95.
\item[-] We impose $p_T>50$~GeV and $|\eta|<2.5$ for each long-lived stau.
\end{itemize}

The trigger efficiency for heavy long-lived sleptons is estimated to be larger than 60\% \cite{ATLAS:longlivedsleptons}. In our calculation we impose this value, in order to be conservative. 
Current searches exclude long-lived staus lighter than $m_{\tilde{\tau}_1}\approx342$~GeV, a bound that we also take into account.

\begin{table}[!t]
\begin{center}
\begin{tabular}{|l|c||c|}
\hline
&\multicolumn{1}{|c||}{$\sqrt{s}=8$ TeV, $\mathcal{L}=20\, \text{fb}^{-1}$}&\multicolumn{1}{|c|}{$\sqrt{s}=13$ TeV, $\mathcal{L}=100\, \text{fb}^{-1}$}\\
&\multicolumn{1}{|c||}{$\sigma^{8\,\text{TeV}}_{\tilde{\chi}_j^\pm\tilde{\chi}_i^0}=1.17\ {\rm fb}$}&\multicolumn{1}{|c|}{$\sigma^{13\,\text{TeV}}_{\tilde{\chi}_j^\pm\tilde{\chi}_i^0}=4.77\ {\rm fb}$}\\
\hline
S3a&1.7&30.3\\
S3b&1.5&28.9\\
\hline
\end{tabular}
\end{center}
\caption{\small Number of events that pass all the cuts for the LHC configurations $\sqrt{s}$ = 8 TeV, $\mathcal{L}=20\, \text{fb}^{-1}$ and $\sqrt{s}$ = 13 TeV, $\mathcal{L}=100\, \text{fb}^{-1}$. An efficiency of 60\% is assumed in the trigger cut.}
      \label{tab:eventss3}
\end{table}

The resulting number of events that pass all the cut is displayed in Table \ref{tab:eventss3} for benchmark points S3a and S3b and considering the current and future LHC configurations. As we observe, none of these benchmark points are observable in the current LHC configuration, since the number of events is below 2 (thereby being in agreement with the negative results of current searches). However, they could be probed in the future with a higher energy and luminosity, for which as many as 30 events could be obtained.

Upon detection, the mass of the stau can be determined using $m_{\tilde{\tau}}=p/\gamma\beta$, where $p$ is the magnitude of the momentum vector of the long-lived particle, $\vec{p}$, and $\beta$ and $\gamma$ are the usual relativistic factors. Notice however that this would not be sufficient to identify this scenario and distinguish it from other possibilities with long-lived charged particles such as the MSSM or NMSSM (when the mass-difference between the stau NLSP and the neutralino LSP is smaller than the tau mass), 
gauge mediated supersymmetry breaking (GMSB) models in which a stau NLSP decays into a tau and a gravitino LSP, or various R-parity breaking models.

A recent analysis of long-lived staus in the MSSM with sneutrinos (which mixed LH-RH states) has been recently presented in Ref.\cite{Cabrera:2013} in which the origin of the long stau lifetime is due to a small mass gap between the LSP and the NLSP. In our case, the stau lifetime is controlled by the small neutrino Yukawa, thereby providing more flexibility in the choice of sparticle masses. In any case, since this signature would be the same, it would be difficult to use it to discriminate between these two scenarios.

\section{Conclusions}
\label{sec:conclusions}

In this work we have investigated exotic collider signatures of the Next-to-Minimal Supersymmetric Standard Model with a right-handed neutrino and sneutrino.
This is a construction in which an extra singlet superfield, $N$, is included in the NMSSM in order to account for RH neutrino and sneutrino states. After electroweak symmetry-breaking takes place, a Majorana mass term is generated for the RH neutrinos which is of the order of the Higgs expectation value and implies an electroweak scale see-saw mechanism, with a small Yukawa coupling $\yn\sim10^{-6}$, for neutrino mass generation.
Such a small neutrino Yukawa leads to a tiny mixing between right and left-handed fields. It is for this reason that the RH neutrino, when produced at the LHC, can be long-lived and give rise to displaced vertices.

We have incorporated the recent constraints on the masses of supersymmetric particles, as well as on low-energy observables. 
We also impose the presence of a Higgs boson with a mass of approximately $125.5$~GeV and consider the existing results on the reduced signal strengths for its decays into Standard Model particles, which place a bound on its invisible and non-standard decays. We study the effect of these constraints on the parameter space of the model.

In the first part of this work we have investigated the production and late decay of RH neutrinos. We show that, due to the small neutrino Yukawa, the RH neutrino can decay in the inner detector of ATLAS or CMS, giving rise to a displaced vertex. This can be observed through the decay products, which involve two leptons ($2\ell + \met$) or a lepton with two jets ($\ell j j $). For a representative number of benchmark points we have simulated the production of RH neutrinos in the current LHC configuration (with a center of mass energy of 8~TeV and an integrated luminosity of $\mathcal{L}=20$ fb$^{-1}$), and a future one (13~TeV and $\mathcal{L}=100$ fb$^{-1}$), defining a number of basic cuts to single out the signal. We have found that some points of the parameter space can already be probed with the current LHC data, and others can become accessible in the future upgrade. We have constructed  the two-lepton ($m_{\ell\ell}$) and two-jets one lepton ($m_{\ell j j }$) invariant mass distributions for the different benchmark points, showing that the end-point in $m_{\ell\ell}$ and the peak in $m_{\ell j j }$ can give valuable complementary information on the mass of the RH neutrino that can help distinguishing this scenario from models with R-parity violation.

In the second part of the analysis we have considered the possibility that the stau is the NLSP. We have shown that the stau decay can also be suppressed by the small Yukawa couplings in certain regions of the parameter space. We have simulated the production of staus in the current and future LHC configuration for two benchmark points. The results suggest that some points in the parameter space can be within the reach of the future LHC configuration.

\vspace*{1cm}
\noindent{\bf \large Acknowledgments}

We thank Chiara Arina, Alexander Belyaev, Mar\'ia Eugenia Cabrera, Camilo Garc\'ia, Pradipta Ghosh, Alejandro Ibarra, Jes\'us Moreno, and Chan-Beom Park for useful discussions.
D.G.C. is supported by the Ram\'on y Cajal program of the Spanish MICINN. 
We also thank the support of the Consolider-Ingenio 2010 programme under grant MULTIDARK CSD2009-00064, the Spanish MICINN under Grant No. FPA2012-34694, the Spanish MINECO ``Centro de excelencia Severo Ochoa Program" under Grant No. SEV-2012-0249, the Community of Madrid under Grant No. HEPHACOS S2009/ESP-1473, and the European Union under the ERC Advanced Grant SPLE under contract ERC-2012-ADG-20120216-320421.


\end{document}